\documentclass[rmp,aps,twocolumn,amsmath,nofootinbib,showpacs]{revtex4}
\usepackage{graphics}
\usepackage{epsfig}
\def\inbar{\,\vrule height1.5ex width.4pt depth0pt}
\def\IR{\relax{\rm I\kern-.18em R}}
\def\IC{\relax\hbox{$\inbar\kern-.3em{\rm C}$}}

\begin{document}
\title{Colloquium: Nonlinear collective interactions in quantum plasmas with degenerate electron fluids}
\author{P. K. Shukla\footnote{E-mail: profshukla@yahoo.de} and B. Eliasson}
\affiliation{RUB International Chair, Fakult\"at f\"ur Physik und Astronomie,
Ruhr-Universit\"at Bochum, D-44780 Bochum, Germany}
\received{27 September 2010; Revised 18 February 2011; Accepted 6 May 2011,
in Reviews of Modern Physics}
\begin{abstract}
The current understanding of some important nonlinear collective processes in quantum plasmas with
degenerate electrons is presented. After reviewing the basic properties of quantum plasmas, we present
model equations (e.g. the quantum hydrodynamic and effective nonlinear Schr\"odinger-Poisson
equations) that describe collective nonlinear phenomena at nanoscales. The effects of the electron
degeneracy arise due to Heisenberg's uncertainty principle and Pauli's exclusion principle for
overlapping electron wavefunctions that result in tunneling of electrons and the electron degeneracy
pressure. Since electrons are Fermions (spin-$1/2$ quantum particles), there also appears
an electron spin current and a spin force acting on electrons due to the Bohr magnetization.
The quantum effects produce new aspects of electrostatic (ES) and electromagnetic (EM) waves 
in a quantum plasma that are summarized in here. Furthermore, we discuss nonlinear features 
of ES ion waves and electron plasma oscillations (ESOs), as well as the trapping of intense
EM waves in quantum electron density cavities. Specifically, simulation studies of the coupled 
nonlinear Schr\"odinger (NLS) and Poisson equations reveal the formation and dynamics of localized
ES structures at nanoscales in a quantum plasma. We also discuss the effect of an external magnetic 
field on the plasma wave spectra and develop quantum magnetohydrodynamic (Q-MHD) equations. 
The results are useful for understanding numerous collective phenomena in quantum plasmas, 
such as those in compact astrophysical objects (e.g. the cores of white dwarf stars and giant planets), 
as well as in plasma-assisted nanotechnology (e.g. quantum diodes, quantum free-electron lasers, 
nanophotonics and nanoplasmonics, metallic nanostructures, thin metal films, semiconductor quantum
wells and quantum dots, etc.), and in the next-generation of intense laser-solid density plasma 
interaction experiments relevant for fast ignition in inertial confinement fusion schemes.
\end{abstract}
\pacs{05.30.Fk, 52.35.Mw, 52.35.Ra, 52.35.Sb}
\maketitle
\tableofcontents

\maketitle

\section{Introduction}

Dense plasmas composed of ions, degenerate electrons, positrons, and/or holes (in the context of
semiconductors) are referred to as quantum plasmas. In the latter, the degeneracy of the lighter
plasma species (electrons, positrons, holes) appears at very high densities and relatively low
temperatures, where the mean inter-particle distance is smaller than (or of the same order as)
the de Broglie thermal wavelength.  The ions are typically non-degenerate due to their relatively
large mass in comparison with the electron mass. In quantum physics, Heisenberg's uncertainty
principle \citep{Dirac81,Holland93,Landau98a,Bransden00} dictates that conjugate variables, such as
the position and momentum of a particle cannot be precisely determined simultaneously; the product
of the uncertainties of the position and momentum is equal to or larger than $\hbar/2$, where
$\hbar ~(=1.0544 \times 10^{-27}~ {\rm erg~ sec}$) is Planck's constant divided by $2\pi$. The position
of an electron subjected to the influence of an atomic nucleus is very well defined (the force to which
it is subjected is large). However, owing to Heisenberg's uncertainty principle, the electron momentum is
ill defined.  An electron has a continuous motion around the position it occupies. This motion exerts pressure
on the surrounding medium, exactly as the thermal agitation of the particles of a gas exerts its pressure.
This pressure is called the electron degeneracy pressure. This pressure, since it is nonthermal in origin,
is, of course, independent of the electron temperature; the pressure of degenerate electrons increases with
increasing electron number density.  It is, however, only at very high densities that the degeneracy pressure
becomes comparable or larger to the thermal gas pressure.  One then says that the plasma matter is in an exotic
state, comprising degenerate electrons and positrons or holes.

Plasmas with degenerate electrons and positrons with number densities comparable with solids and temperatures
of several electron volts fall under the category of dense matter \citep{Ichimaru82,Fortov09} that appears in
the core of giant planets \citep{Chabrier06,Chabrier09,Horn91} and the crusts of old stars \citep{Guillot99}.
Dense compressed plasmas are currently of wide interest due to their applications to astrophysical and
cosmological environments \citep{Lai01,Harding06,Opher01,Benvenuto05}, as well as to inertial fusion science
involving intense laser-solid density plasma interaction
experiments \citep{Lindl95,Hu99,Andreev00,Mendonca01,Son05,Salamin06,Marklund06, Malkin07,Glenzer07,Kritcher08,
Glenzer09,Lee09,Neumayer10,Froula11} for inertial confinement fusion \citep{Azechi06} based on the high-energy
density plasma physics \citep{Drake09,Drake10,Norreys09}. Plasma-like collective behavior is well studied 
experimentally and theoretically in solid state physics \citep{Kittel96}, in which metals and semi-conductors 
support both transverse optical modes, and longitudinal electrostatic modes, such as plasmons and phonons on 
electron and ion time-scales, and, in addition various lattice modes. Plasmons and phonons are usually probed by
measuring the energy of electrons which have been passed through thin foils, or by laser scattering techniques.
For example, the dispersion relation of collective electron plasma waves has been measured for several metal
specimen by using an electron velocity analyzer of M\"ollenstedt type \cite{Watanabe56}. Collective dispersive
behavior of plasmons, including shifts in the plasmon frequency due to quantum effects, in solid-density plasmas
have been observed by \citet{Glenzer07} and \citet{Neumayer10} using spectrally resolved x-ray scattering
techniques \citep{Kritcher08,Lee09}.  In these experiments, powerful x-ray sources are employed for
accessing narrow bandwidth spectral lines via collective Thomson scattering of light off electron-density
fluctuations. These experimental techniques also allow accurate measurements of the electron velocity
distribution function, temperature, and ionization state in the dense matter regime. \citet{Gregori09}
also proposed future experiments to measure low-frequency oscillations in plasmas when keV free-electron
lasers will become available.  \citet{Froula11} summarizes the measurement techniques using scattering of
EM waves in plasmas, and recent experimental results from x-ray scattering experiments in dense plasmas
reveal that quantum mechanical effects are indeed important \citep{Glenzer07,Glenzer09}.

Furthermore, due to recent experimental progress in femtosecond pump-probe spectroscopy, the field of quantum
plasmas is also gaining significant attention \citep{Crouseilles08} in connection with the collective dynamics
of an ensemble of degenerate electrons in metallic nanostructures and thin metal films. The physics of quantum
plasmas is also relevant in the context of quantum diodes \citep{Ang03,Ang07,Shukla08b}, nanophotonics and
nanowires \citep{Barnes03,Shpata06,Chang06}, nanoplasmonics \citep{Ozbay06,Maier07,Atwater07,Marklund08,Stockman11},
high-gain quantum free-electron lasers \citep{Serbeto08,Serbeto09}, microplasma systems \citep{Becker06},
and small semiconductor devices \citep{Markowich90,Haug04,Haug07,Manfredi07}, such as quantum wells and
piezomagnetic quantum dots \citep{Adolfath08}. The latter can be used as nanoscale magnetic switches.

Collective interactions between an ensemble of degenerate electrons and positrons/holes give rise
to novel waves and structures in quantum plasmas. Studies of linear waves in a non-relativistic
unmagnetized quantum plasma with degenerate electrons begun with the pioneering theoretical
works of \citet{Silin52a,Silin52b,Silin61,Bohm1,Bohm2} and \citet{Pines61}, who studied the dispersion
properties of high-frequency electron plasma oscillations (EPOs). The frequencies of the latter with
an arbitrary electron degeneracy have been found by \citet{Maafa93}. In the theoretical description of
the EPOs, Klimontovich-Silin and Bohm-Pines used the Wigner distribution function \citep{Wigner32}
and the density matrix approach to demonstrate that in a quantum plasma with a Fermi-Dirac equilibrium
distribution function for degenerate electrons, the frequency of the EPOs is significantly different from
the Bohm-Gross frequency in a classical electron-ion plasma with non-degenerate electrons obeying the
Maxwell-Boltzmann distribution function. The dispersion to the EPOs appears through the electron Fermi
pressure and electron tunneling effects \citep{Wilhelm71,Gardner96,Manfredi01,Manfredi05,
Shukla06,ShuklaEliasson06,Jungel06,Misra09,Shukla10}. The quantum Bohm potential, responsible for electron
tunneling, appeared first in the quantum fluid description of a single electron by \citet{Madelung26} 
and \citet{Bohm52}. For systems of degenerate electrons, different forms of the potential have been 
derived by using moments of the Wigner equation \cite{Iafrate81,Ancona89} and by using a variational 
approach \citep{Feynman86,Kleinert86}. They have been used in quantum hydrodynamic (QHD) 
equations \cite{Wilhelm71} for modeling nano-devices \citep{Ferry93,Gardner96}. More recently, 
Lagrangian approaches have been used to device efficient computational algorithms for quantum 
systems \citep{Mayor99,Lopreore99}. These and other methods for computational QHD using quantum 
trajectories have been nicely summarized in the textbook by \citet{Wyatt05}.

The quantum effects are important for the dielectric and dispersive properties of a quantum plasma.
The longitudinal and transverse dielectric constants of an isotropic quantum plasma were worked 
out by \citet{Lindhard54}, \citet{Silin61a},and \citet{Kuzelev99}.  Contributions of the electron 
spin and exchange interactions to the electromagnetic (EM) wave dispersion relations in an unmagnetized 
quantum plasma have been presented by \citet{Burt62} by using a quantum kinetic theory. The quantum 
mechanical phase space distribution of \citet{Wigner32} has been further generalized by \citet{Brittin62}
for a system of charged particles including the quantized EM field and Green's functions involving 
correlations of distribution functions and vector potentials.  Kinetic models for spin-polarized plasmas 
have been developed by \citet{Cowley86}, \citet{Zhang88}, and \citet{Balescu88}. More recently, electron 
spin-1/2 effects in a quantum magnetoplasma have also been considered by \citet{Brodin08a} and \citet{Zamanian10}. 
The gauge problem in quantum kinetics has been treated by \citet{Stratonovich56} and \citet{Serimaa86}, 
which is important whenever the fields are not electrostatic. In a quantum magnetoplasma, one finds that 
the external magnetic field significantly affects the dynamics of degenerate electrons, and that the 
thermodynamics and kinetics \cite{Steinberg00} in a quantum magnetoplasma are significantly different 
from those in an unmagnetized quantum plasma.  \citet{Oberman63} derived the expression for the dielectric 
function for longitudinal waves in a non-relativistic magnetized quantum plasma and discussed applications 
of their work to heavily doped semiconductors. \citet{Kelly64} studied the dispersive properties of a 
magnetized quantum plasma by using the Wigner distribution function and the Maxwell equations.  
Finally, we mention that useful foundations for the theory of quantum plasmas are presented 
by \citet{deGroot72}, while quantum kinetic models including the effects of spin are reviewed by \citet{Lee95}.

During the last decade, there has been a surge in investigating new aspects of collective interactions
in dense quantum plasmas by means of non-relativistic quantum hydrodynamic
\citep{Gardner96,Manfredi01,Manfredi05,Jungel06,Shukla10} and quantum kinetic \citep{Bonitz98,Kremp99,Tsintsadze09}
equations. Models for non-ideal effects in a strongly coupled dense plasma have been presented 
by \citet{Carruthers83}, \citet{Kremp05}, and \citet{Redmer10}. The Wigner-Poisson (WP) 
model \citep{Hillery84} has been used to derive a set of QHD equations \cite{Manfredi01,Manfredi05} 
for ES waves in a quantum plasma. The relation between the QHD and kinetic models have been investigated 
by \citet{Haas10b}. The quantum nature \citep{Manfredi01,Shukla10} is manifested in the non-relativistic electron
momentum equation through the quantum statistical pressure, which requires the knowledge of the Wigner electron
distribution function for a quantum mixture of electron wavefunctions, each characterized by an occupation
probability. The quantum part of the electron pressure is also represented as a nonlinear quantum
force \citep{Wilhelm71,Gardner96,Manfredi01} $-\nabla \phi_B$, where
$\phi_B = - (\hbar^2/2 m_e \sqrt{n_e})\nabla^2 \sqrt{n_e}$ is the Bohm potential, and $m_e$ and
$n_e$ are the electron mass and electron number density, respectively.. Defining the effective
wavefunction $\psi =\sqrt{n_e({\bf r},t)} \exp[iS_e({\bf r}, t)/\hbar]$, where $\nabla S_e({\bf r}, t)
= m_e {\bf u}_e({\bf r},t)$ and ${\bf u}_e ({\bf r}, t)$ is the electron fluid velocity, the non-relativistic
electron momentum equation can be cast into an effective nonlinear Schr\"odinger (NLS) equation
\cite{Manfredi01,Manfredi05,Shukla06,ShuklaEliasson06,Shukla10}, in which there appears a coupling between
the electron wavefunction and the ES potential associated with the EPOs. The ES potential, in turn,
is determined from Poisson's equation. One thus has the coupled NLS and Poisson equations, governing
the dynamics of nonlinearly interacting EPOs is a quantum plasma.  Both non-relativistic QHD and NLS-Poisson
equations exclude strong interactions among the quantum particles and electron exchange
interactions \citep{Hohenberg64,Kohn65} between an electron and the background plasma particles
(e.g. degenerate electrons and non-degenerate ions).  However, it has turned out that the QHD and
NLS-Poisson equations have been quite useful for studying linear and nonlinear plasma waves, as well
as stability of quantum plasmas \citep{Manfredi01,Manfredi05,Shukla06,ShuklaEliasson06,Shukla10,Haas03,Haas05,Haas07} 
at nanoscales involving the quantum force \citep{Wilhelm71,Gardner96} and the quantum statistical pressure
law for an unmagnetized quantum plasma with degenerate electrons. New effects also appear when one accounts 
for the  potential energy of the electron spin--$1/2$ in a magnetic 
field \citep{Takabayasi55,Marklund07,Misra07,Brodin07a,Brodin07b,Brodin07c,Brodin08b,Shukla2007,Shukla09,
Misra09,Misra10,Brodin10}. In fact, the QHD model for degenerate electrons in both
non-relativistic \citep{Manfredi01,Manfredi05,Shukla06,ShuklaEliasson06,Shukla10} and relativistic
\citep{Masood10} quantum plasma regimes seems to provide an adequate description for probing some quantum
collective interactions in compressed plasmas \citep{Glenzer07,Glenzer09,Lee09,Neumayer10,Froula11} due
to the availability of ultrafast x-ray Thompson scattering spectroscopic techniques.

In this Colloquium, we present the recent development of numerous nonlinear collective processes in a quantum
plasma with degenerate electrons. We first describe the salient properties of quantum plasmas in which
degenerate electrons follow the Fermi-Dirac distribution. We then present the relevant equations for describing 
linear and nonlinear wave phenomena in quantum plasmas. After reviewing the linear properties of ES and EM waves, 
we proceed by presenting numerical results of the governing nonlinear equations which reveal localization of 
ES and EM waves at nanoscales. Specifically, we discuss the formation and dynamics of nanostructures 
(e.g. 1D quantum electron density cavity and 2D quantum vortices), as well as discuss the properties of 
3D quantum electron fluid turbulence at nanoscales. Also presented are nonlinear interactions between intense 
EM waves and ESOs, which reveal stimulated scattering of EM waves off quantum plasma oscillations and 
trapping of light into a quantum electron density cavity. The effects of an external magnetic field 
on linear and nonlinear wave phenomena in a quantum magnetoplasma are examined. Finally, we highlight
possible applications, as well as future perspectives and outlook of nonlinear quantum plasma physics.

\section{Basic Properties of Quantum Plasmas}

Let us first summarize some of the basic properties of quantum plasmas that are quite distinct
from classical plasmas. While classical plasmas are composed of non-degenerate plasma particles
with low number densities and relatively high electron and ion temperatures, quantum plasmas
have degenerate electrons and/or positrons with extremely high number densities and relatively low
temperatures. The ions can usually be treated as non-degenerate plasma particles. Figure 1 depicts 
the plasma parameter regimes (the electron temperature versus the electron number density) under 
which quantum plasmas occur in different physical environments.

\begin{figure}[htb]
\centering
\includegraphics[width=8cm]{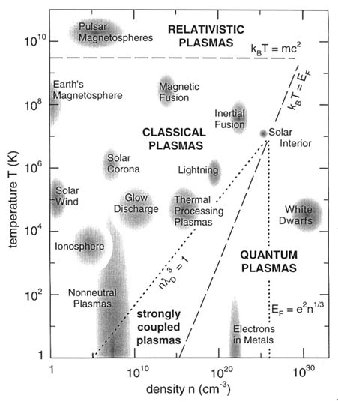}
\caption{The plasma diagram in the log $T$- log $n_e$ plane, separating the classical and
quantum regimes. After \citet{Plasma95}.}
\label{Fig0}
\end{figure}

Quantum mechanical effects start playing a significant role when the Wigner-Seitz radius (average 
inter-particle distance) $a = (3/4\pi n)^{1/3}$ is comparable to or smaller than the thermal de~Broglie 
wavelength $ \lambda_B = \hbar/m V_T$, where $m$ is the mass of the quantum particles (e.g. degenerate 
electrons, degenerate positrons, degenerate holes), $V_T = (k_B T/m)^{1/2}$ is the thermal speed of the
quantum particles, $T$ is the temperature, $m$ is the mass, and $k_B$ is the Boltzmann constant, i.e. when
\begin{equation}
n \lambda_B^3 \geq 1,
\end{equation}
or, equivalently, when the temperature $T$ is comparable to or lower than the Fermi temperature $T_F= E_F/k_B$,
where the Fermi energy is
\begin{equation}
E_F =\frac{\hbar^2}{2m} (3\pi^2)^{2/3}n^{2/3}.
\end{equation}
The relevant degeneracy parameter for the quantum plasma is
\begin{equation}
\frac{T_F}{T} = \frac{1}{2}(3\pi^2)^{2/3}({n \lambda_B^3})^{2/3} \geq 1.
\end{equation}
For typical metallic densities of free electrons, $n \sim 5 \times 10^{22}$ cm$^{-3}$, we have
$T_F \sim 6\times 10^4\,\mathrm{K}$, which should be compared with the usual temperature $T$.

When the plasma particle temperature approaches $T_F$, one can show, by using a density matrix
formalism \citep{Bransden00}, that the equilibrium distribution function changes from the
Maxwell--Boltzmann $\propto \exp(-E/k_B T)$ to the Fermi--Dirac (FD) distribution function
\begin{equation}
{\cal F}_{FD} = 2\left(\frac{m}{2\pi \hbar}\right)^{3}
\left[ 1 + \exp\left(\frac{E -\mu}{k_B T}\right)\right]^{-1},
\end{equation}
where in the non-relativistic limit the energy is $E = (m/2) v^2 = (m/2) (v_x^2 + v_y^2 + v_z^2)$.
The chemical potential is denoted by $\mu$.  The parameter $\mu/k_B T$ is large and negative in the
non-degenerate limit, and is large and positive in the completely degenerate limit. The equilibrium
electron number density associated with the FD distribution function is
\begin{equation}
n_0 =\int {\cal F}_{FD}\,d^3v = - \frac{1}{4}\left(\frac{2m k_B T}{\pi \hbar^2}\right)^{3/2}
{\rm Li}_{3/2}[-\exp(\xi_\mu)],
\end{equation}
where  ${\rm Li}_{3/2}$ is the poly-logarithm function, and $\xi_\mu =\mu /k_B T$. The completely
degenerate limit corresponds to $\mu \rightarrow k_B T_F$ and $T_F \gg T$. The relation between 
$T_F/T$ and $\xi_\mu$ is \citep{Melrose08} $-Li_{3/2}[- \exp(\xi_\mu)] =(4/3\sqrt{\pi})(T_F/T)^{3/2}$.

It us useful to define the quantum coupling parameters for electron-electron and ion-ion interactions.
The electron-electron Coulomb coupling parameter is defined as the ratio between the electrostatic
interaction energy $E_{int} =e^2/a_e$ between electrons and the electron Fermi energy $E_{Fe} =k_B T_{Fe}$,
where $e$ is the magnitude of the electron charge and $a_e =(3/4\pi n_e)^{1/3}$ is the mean inter-electron
distance. We have
\begin{equation}
\Gamma_{e} = \frac{E_{int}}{E_{Fe}} \approx 0.3 \left(\frac{1}{n_e \lambda_{Fe}^3}\right)^{2/3} \approx
0.3 \left(\frac{\hbar \omega_{pe}}{k_B T_{Fe}}\right)^2,
\end{equation}
where $\lambda_{Fe} = V_{Fe}/\omega_{pe}$, $V_{Fe} =(2E_{Fe}/m_e)^{1/2}
= (\hbar/m_e)(3\pi^2 n_e)^{1/3}$ is the electron Fermi speed, and $\omega_{pe} = (4\pi n e^2/m_e)^{1/2}$
the electron plasma frequency. Furthermore, the ion-ion Coulomb coupling parameter is
$\Gamma_i =Z_i^2 e^2/a_i k_B T_i$, where $Z_i$ is the ion charge state, $a_i= (3/4\pi n_i)^{1/3}$
the mean inter-ion distance, and $T_i$ the ion temperature.

Since $\Gamma_e$ for metallic plasmas could be larger than unity, it is of interest to enquire the role of
inter-particle collisions on collective processes in a quantum plasma. It turns that the Pauli blocking
reduces the collision rate for most practical purposes \citep{Manfredi05,Son05}. Due to
Pauli blocking, only electrons with a shell of thickness $k_B T$ about the Fermi surface suffer collisions.
For these electrons, the electron-electron collision frequency is proportional to $k_B T/\hbar$. The average
collision frequency among all electrons turns out to be \citep{Manfredi05}
\begin{equation}
\nu_{ee} = \frac{k_B T^2}{\hbar T_{Fe}}.
\end{equation}
Typically, $\nu_{ee} \ll \omega_{pe}$ when $T < T_{Fe}$, which is relevant for metallic electrons.
On the other hand, the typical timescale for electron-ion (lattice) collisions is $\tau_{ei} \simeq 10$ fs,
which is one order of magnitude greater than the electron plasma period. Accordingly, a collisionless
quantum plasma regime is relevant for phenomena appearing on the timescale of the order of a femtosecond
in a metallic plasma.

In compact astrophysical objects such as white dwarf stars, the mean distance $n_e^{-1/3}$ between
electrons become comparable to the Compton length $\lambda_C=\hbar/m_e c$, and accordingly the speed
of an electron on the Fermi surface becomes comparable to the speed of light $c$ in vacuum, so that one
has to take relativistic effects into account. Relativistic degenerate electrons are found in the core 
of massive white dwarf stars \citep{Shapiro83,Koester90}, aptly named due to their very low luminosities 
yet high surface emissivities, which are compact bodies with radii $\leq 10^{-2} ~R_{\odot}$ and masses 
typically $\leq M_{\odot}$.  Consequently, the average electron number densities are quite high 
($\sim 10^{30}$ cm$^{-3}$). Since electrons are Fermions, only one electron can occupy a given quantum 
state (position, spin). In a simplified picture, each electron will on average occupy a volume $1/n_e$.  
Then, by Heisenberg's uncertainty principle \citep{Bransden00}, $\Delta x \Delta p \lesssim \hbar/2$, 
the mean momentum of electrons can be estimated to be $p_x \approx \hbar n_e^{1/3}$. If electrons are 
non-relativistic, the velocity of the electron is $\sim  p_x/m_e =\hbar n_e^{1/3}/m_e$; however, if electrons 
are relativistic, their velocity will be close to $c$.  Now the electron pressure, as it is for a simple gas,
is the momentum transfer per unit area, or $P_e =$ (momentum) $ \times$ (velocity) $\times$ (number density).
For non-relativistic electrons, we have \citep{Gursky76} $P_e = \hbar n_{e}^{1/3}(\hbar n_{e}^{1/3}/m_e) n_e
= \hbar^2 n_{e}^{5/3}/m_e$.  On the other hand, when electrons are relativistic, the relativistic electron pressure
is $P_{er} =\hbar n_{e}^{1/3} c n_e =\hbar c n_{e}^{4/3}$. In the 
past, \citet{Chandra31a,Chandra31b,Chandra35,Chandra39} and others presented a rigorous derivation of 
the electron pressure $P_C$ for arbitrary relativistic electron degeneracy pressure in dense matter. It reads
\begin{equation}
P_C = \frac{\pi}{3 h^3} m_e^4 c^5 f(\xi_c),
\end{equation}
where $f(\xi_c) =\xi_c(2\xi_c^2-3)(1+\xi_c^2)^{1/2} + 3 {\rm sinh}^{-1} (\xi_c)$, $\xi_c =p_ce/m_ec$,
and $p_c =(3h^3n_e/8\pi)^{1/3}$ is the momentum of an electron on the Fermi surface. In the
non-relativistic limit $\xi_c \ll 1$, we have \citep{Chandra35,Chandra39}
\begin{equation}
P_n =  \frac{(\pi)^{2/3}}{5 m_e} \hbar^2 n_e^{5/3},
\end{equation}
while in the ultra-relativistic limit $\xi_c \gg 1$, the degenerate electron pressure reads \citep{Chandra31a}
\begin{equation}
P_u =  \frac{\left(3\pi^2 \right)^{1/3}}{4} \hbar c n_e^{4/3} = \frac{3}{4}\hbar c n_e^{4/3}.
\end{equation}
Thus, the intuitively obtained formulas of \citet{Gursky76} for non-relativistic and ultra-relativistic
pressures for degenerate electrons are in agreement with those deduced from the pressure formula (8) 
for an arbitrary relativistic electron degeneracy pressure.

In his Nobel Prize winning papers on the structure of compact stars, \citet{Chandra31a,Chandra31b} balanced 
the gradient of the ultra-relativistic electron degeneracy pressure $P_u/R$ and the gravitational force
$G =(GM/R^2) n_e m_n$, where $G$ is the gravitational constant, $M$ and $R$ are the mass and radius
of a star, respectively, $m_n$ the mass of the nuclei $(n_e m_n =M /R^3)$, to deduce the critical mass
of a star $M_c =\left(\hbar c/G\right)^{3/2}m_n^{-2} \approx 1.4 ~M_{\odot}$, where $M_{\odot}$ is the
solar mass.  The interior of white dwarf stars usually consists of fully ionized helium, carbon, and oxygen,
which approximately consist of equal amounts of protons and neutrons. Hence, the effective mass of the
nuclei can be taken to be the proton mass plus the neutron mass. Since $M_c$ is independent of density,
it means that this mass is obtained independent of radius. This is the limiting mass; more massive
stars cannot be supported by electron degeneracy pressure no matter how small they are. This was
the discovery of Chandrasekhar; that the pressure dependence on density changed in going from
nonrelativistic to relativistic conditions and, as a consequence, there arose a finite limit to
the mass of a star with ultra-relativistic degenerate electrons. 

\section{Model Equations for Quantum Plasmas}

In quantum systems, the Dirac and Maxwell equations are often used to study the dynamics of
a relativistic quantum particle/Fermion (electrons and positrons) in the presence
of intense electromagnetic fields. Quantum particles have spin.  For example, an electron spin
$s=1/2$ is an intrinsic property of electrons which have an intrinsic angular momentum
characterized by quantum number $1/2$, and a magnetic moment for individual electrons.
In fact, the relativistic Dirac equation provides a description of quantum particles (with spin) under
the action of the electromagnetic fields. The spin of electrons (and positrons)-which have the spin-$1/2$
has been introduced through Dirac's Hamiltonian \citep{Dirac81}

\begin{equation}
{\cal H} = c {\boldsymbol \alpha}_s \cdot ({\bf p}_e + \frac{e}{c} {\bf A})
- e \phi + {\boldsymbol \beta} m_e c^2,
\end{equation}
where ${\bf p}_e = - i\hbar \nabla$ is the momentum operator, and ${\boldsymbol \alpha}_s$ and $\boldsymbol{\beta}$
are the Dirac matrices. The three Cartesian components $\alpha_j$ ($j=1,2,3$) of ${\boldsymbol \alpha}_s$ are
usually constructed with help of the Pauli spin matrices $\sigma_x$, $\sigma_y$ and $\sigma_z$ \citep{Bransden00}.
The corresponding wave functions $\psi$ are four-component spinors. The magnetic field is ${\bf B} =
\nabla \times {\bf A}$, where ${\bf A}$ and $\phi$ are the vector and scalar potentials, 
which are determined from the Maxwell equations.

In the non-relativistic limit, the Pauli equation \citep{Pauli99} in the presence of the electromagnetic fields 
describes the dynamics of a single quantum particle. It reads \citep{Tsintsadze09}
\begin{equation}
i \hbar \frac{\partial \psi_\alpha}{\partial t} = H_\alpha \psi_\alpha,
\end{equation}
where
\begin{equation}
\begin{split}
H_\alpha = & - \frac{\hbar^2}{2m_\alpha} \nabla^2  - \frac{i q_\alpha \hbar}{2m_\alpha c}\left({\bf A} \cdot \nabla
+ \nabla \cdot {\bf A}\right) 
\\
&+ \frac{q_\alpha^2 {\bf A}^2 }{2m_\alpha c^2} + q_\alpha \phi
- {\bf \mu}_\alpha \cdot {\bf B},
\end{split}
\end{equation}
is the Hamiltonian, and $\psi_\alpha ({\bf r}, t, \boldsymbol{\sigma})$ is the wavefunction of the single 
quantum particle species $\alpha$ with the spin ${\bf s} = (1/2) \boldsymbol{\sigma}$ ($|\boldsymbol{\sigma}| = 1)$, 
and $q_\alpha = -e$ $(+e)$ for electrons (positrons). The last term in (13) is the potential energy of the
magnetic dipole in the external magnetic field, the magnetic moment of which is ${\bf \mu}_\alpha 
= (q_\alpha \hbar/2m_\alpha c)\boldsymbol{\sigma} \equiv \mu_B \boldsymbol{\sigma}$, where 
$\mu_B =q_\alpha\hbar/2 m_e c$ is the Bohr-Pauli magneton and $\boldsymbol{\sigma}$ the spin-operator 
of a single quantum particle \citep{Landau98a}.

By using the Madelung representation \citep{Madelung26} for the complex wavefunction $\psi_\alpha$, viz.
\begin{equation}
\psi_\alpha ({\bf r}, t, {\bf \sigma}) = \Psi_\alpha({\bf r}, t, {\bf \sigma})\exp\left(i S_\alpha/\hbar\right),
\end{equation}
where $\Psi_\alpha ({\bf r}, t, {\sigma})$ and $S_\alpha({\bf r}, t, {\bf \sigma})$ are real, in the Pauli
equation (12), we obtain the quantum Madelung fluid equations \citep{Tsintsadze09}
\begin{equation}
\frac{\partial n_\alpha}{\partial t} + \nabla \cdot \left(n_\alpha {\bf p}_\alpha/m_\alpha\right) =0,
\end{equation}
and
\begin{equation}
\frac{d{\bf p}_\alpha}{dt} = q_\alpha\left({\bf E} + \frac{{\bf u}_\alpha \times {\bf B}}{c}\right)
+ {\bf F}_Q + {\bf F}_s,
\end{equation}
where we have denoted
\begin{equation}
{\bf F}_Q = \frac{\hbar^2}{2m_\alpha} \nabla \left(\frac{\nabla^2 \sqrt{n_\alpha}}{\sqrt{n_{\alpha}}}\right),
\end{equation}
and
\begin{equation}
{\bf F}_s = \mu_B \nabla \left({\bf \sigma}\cdot {\bf B}\right).
\end{equation}
Here $n_\alpha =|\Psi_\alpha|^2$ is the probability density of finding a single quantum particle
with a spin ${\bf s}$ at some point in space, ${\bf p}_\alpha = \nabla S_\alpha -q_\alpha {\bf A}/c$
is the momentum operator of a quantum particle, $d/dt =(\partial/\partial t) + {\bf u}_\alpha \cdot
\nabla$, ${\bf u}_\alpha$ is the velocity of a quantum particle, and ${\bf E} =-\nabla \phi -c^{-1}
\partial {\bf A}/\partial t$ and ${\bf B} = \nabla \times {\bf A}$.

The spin force ${\bf F}_s$ in a quantum magnetoplasma can also be written
as \citep{Marklund07,Brodin07a,Brodin07b,Brodin07c,Brodin08b}
\begin{equation}
{\bf F}_s = \mu_B \tanh \left(\frac{\mu_B B}{k_B T_\alpha}\right) \nabla B,
\end{equation}
where $B = |{\bf B}|$ and $\tanh (\xi) =B_{1/2} (\xi)$, with the Brillouin function
with argument $"1/2"$ describing particles of spin-$1/2$. The Langevin parameter $\tanh(\xi)$
accounts for the macroscopic magnetization of electrons due to the electron thermal agitation
and electron-electron collisions.

\subsection{The Schr\"odinger and Wigner-Poisson Equations}

The quantum $N$-body problem is governed by the Schr\"odinger equation for the $N$-particle
wavefunction $\psi(q_1,q_2,\ldots, q_N)$, where $q_j=({\bf r}_j,s_j)$ is the coordinate
(space, spin) of the particle $j$, each particle associated with energy ${\cal E}_j$.
A drastic simplification occurs if one neglects the correlation between the particles at every
order in $\Gamma_Q$ and describes the full wavefunction as the product of the single particle
wavefunctions. For identical quantum particles, the $N$-particle wavefunction is given by
the Slater determinant \citep{Bransden00}
\begin{equation}
\begin{split}
&\psi(q_1,\,q_2,\,\ldots,\,q_N) 
\\
&= \frac{1}{\sqrt{N!}}
\left| \begin{array}{cccc}
\psi_1(q_1,t)& \psi_2(q_1)& \cdots &\psi_N(q_1)\\
\psi_1(q_2,t)& \psi_2(q_2)& \cdots &\psi_N(q_2)\\
\vdots & \vdots& \ddots & \vdots\\
\psi_1(q_N)& \psi_2(q_N)& \cdots &\psi_N(q_N)\\
\end{array}\right|,
\end{split}
\end{equation}
which is {\sl antisymmetric} with respect to an interchange of any two particle coordinates. This property
is required by the Pauli exclusion principle under the second quantization procedure for a system of
$N$ identical non-relativistic quantum particles. Accordingly, $\psi$ vanishes if two rows are identical,
i.e. two identical quantum particles cannot occupy the same state.  Example $(N=2)$:
$\psi(q_1,q_2)=\frac{1}{\sqrt{2}}[\psi_1(q_1)\psi_2(q_2)-\psi_1(q_2)\psi_2(q_1)]$
so that $\psi(q_2,q_1)=-\psi(q_1,q_2)$ and $\psi(q_1,q_1)=0$. In the zero temperature
limit, all energy states up to the Fermi energy level are occupied, while no energy
states above the Fermi level are occupied.

To capture collective effects in quantum plasmas, \citet{Haas00} and \citet{Anderson02}
used the time-dependent Hartree model where electrons are described by a statistical mixture
of $N$ pure states, where each wavefunction $\psi_j$, $j=1,\,\ldots,\, N$ obeys the
Schr\"odinger equation \citep{Anderson02}
\begin{equation}
i\hbar\frac{\partial \psi_j}{\partial t}+\frac{\hbar^2}{2 m_e}\nabla^2\psi_j+e\phi \psi_j=0,
\end{equation}
which is coupled with Poisson's equation
\begin{equation}
\nabla^2 \phi = 4 \pi e \left(\sum_{j=1}^N |\psi_j|^2 - Z_i n_i\right),
\end{equation}
where $n_i$ is the ion number density (to be obtained from the hydrodynamic equations for non-degenerate ions, 
to be discussed later), and $\phi$ the electrostatic potential arising from the charge distribution of $N$ 
electrons. Equations (21) and (22) have been used to study streaming instabilities \citep{Anderson02} and 
other kinetic effects in a quantum system composed of an ensemble of electrons. Within the Hartree-Fock model,
Eq. (21) can be further generalized by including the electron-exchange term resulting from the Pauli exclusion 
principle. The effect of exchange is for electrons of like-spin to avoid each other. Each electron of a given 
spin is consequently surrounded by an "exchange hole", a small volume around the electron which like-spin 
electrons avoid. For the study of magnetic ordering in  quantum dots (QDs) doped with magnetic impurities, 
Eqs. (21) and (22) must also be enlarged by including a 3D QD confining potential and a Vosko-Wilk-Nusair 
spin dependent exchange-correlation potential \citep{Dharma95}. Hence, the self-consistent model will go far
beyond the Kohn-Sham's description \citep{Kohn65} for treating the dynamics of correlated electrons in
electron clusters, accounting for electron-exchange and electron-correlation effects. In the
presence of time-dependent potentials, the properties and dynamics of many-electron systems can be
investigated by using a time-dependent functional theory \citep{Runge84}.

However, in a non-relativistic quantum plasma with an ensemble of degenerate electrons, it is more 
appropriate to use the quantum statistical theory involving the Wigner distribution 
function \citep{Wigner32}

\begin{equation}
\begin{split}
f_w&({\bf r}, {\bf v}) =\left(\frac{m_e}{2\pi \hbar}\right)^3
\\
&\times \!\! \int \exp(i m_e {\bf v} \cdot {\bf R}/\hbar)
\psi^\ast({\bf r} + {\bf R}/2) \psi({\bf r} -{\bf R}/2)\, d^3 R,
\end{split}
\end{equation}
where the asterisk denotes the complex conjugate. Equation (23) has also been used by Moyal \citep{Moyal49}
for studying the dynamics of electrons in a quantum system. 

For electrostatic interactions in a quantum plasma, the Wigner-Poisson equations, to a leading order 
(in the limit of weak quantum coupling parameter $\Gamma_e$), can be written as
\begin{equation}
\begin{split}
&\frac{\partial f_w}{\partial t} + {\bf v}\cdot \nabla f_w
= -\frac{iem_e^3}{(2\pi)^3  \hbar^4} \int\!\!\!\int e^{im_e({\bf v}-\bf{v}')\cdot {\bf R}/\hbar}
\\
&\times\bigg[\phi\bigg({\bf x }+\frac{\bf R}{2},t\bigg)-\phi\bigg({\bf x}
-\frac{\bf R}{2},t\bigg)\bigg] f_w({\bf x},{\bf v}',t)\, d^3 R \, d^3v'
\end{split}
\end{equation}
and
\begin{equation}
\nabla^2 \phi =4 \pi e\left(\int f_w d^3v - Z_i n_i \right).
\end{equation}

\subsection{The QHD Equations}

The non-relativistic QHD equations \citep{Wilhelm71} have been developed in condensed matter
physics \cite{Gardner96} and in plasma physics \citep{Manfredi01,Manfredi05}. The non-relativistic
QHD equations are composed of the electron continuity equation
\begin{equation}
\frac{\partial n_e}{\partial t} + \nabla \cdot (n_e {\bf u}_e) =0,
\end{equation}
the electron momentum equation \cite{Wilhelm71}
\begin{equation}
m_e\left(\frac{\partial {\bf u}_e}{\partial t} + {\bf u}_e\cdot \nabla {\bf u}_e\right)
= e\nabla \phi -\frac{1}{n_e} \nabla P_e + {\bf F}_Q,
\end{equation}
and Poisson's equation
\begin{equation}
\nabla^2\phi= 4\pi e (n_e -Z_i n_i).
\end{equation}

In a quantum plasma with non-relativistic degenerate electrons, the quantum statistical pressure 
in the zero electron temperature limit can be modeled as \citep{Manfredi01,Crouseilles08}
\begin{equation}
P_e = \frac{m_e V_{Fe}^2n_0}{3}\left(\frac{n_e}{n_0}\right)^{(D +2)/D},
\end{equation}
where $D$ is the number of space dimension of the system, and $V_{Fe} = (\hbar/m_e)(3\pi^2 n_e)^{1/3}$
the electron Fermi speed.

\subsection{The NLS-Poisson Equations}

For investigating nonlinear properties of dense quantum plasmas, it is appropriate to work with
a NLS equation. Hence, by introducing the wavefunction
\begin{equation}
\psi ({\bf r}, t) =\sqrt{n_e ({\bf r}, t)}\exp\left( i\frac{S_e ({\bf r}, t)}{\hbar}\right),
\end{equation}
where $S_e$ is defined according to $m_e {\bf u}_e = \nabla S_e$ and $n_e = |\psi|^2$, it can be
shown that (27) can be cast into a NLS equation \citep{Manfredi01,Manfredi05}
\begin{equation}
i \hbar \frac{\partial \psi}{\partial t} + \frac{\hbar^2}{2m_e}\nabla^2 \psi + e \phi \psi
- \frac{m_e V_{Fe}^2}{2n_0^2}|\psi|^{4/D} \psi =0,
\label{NLS}
\end{equation}
where the electrostatic field $\phi$ is determined from Poisson's equation
\begin{equation}
  \nabla^2 \phi = 4\pi e(|\psi^2| - Z_i n_i).
  \label{Poisson}
\end{equation}
We note that the third and fourth terms in the left-hand side of Eq. (\ref{NLS}) represent the 
nonlinearities associated with the nonlinear coupling between the electrostatic potential and the
electron wavefunction and the nonlinear quantum statistical pressure, respectively.

\section{Linear Waves in Quantum Plasmas}

\subsection{Electron Plasma Oscillations (EPOs)}

Linearization of the NLS-Poisson Equations (31) and (32) around the equilibrium state and combining
the resultant equations, we  obtain the frequency $\omega$ of the EPOs \citep{Silin52a,Silin52b,Bohm1,Bohm2}
\begin{equation}
\omega = \left(\omega_{pe}^2 +  \frac{3}{5} k^2 V_{Fe}^2 + \frac{\hbar^2k^4}{4m_e^2}\right)^{1/2},
\end{equation}
where $k$ is the wavenumber and $\omega_{pe} =(4\pi n_0 e^2/m_e)^{1/2}$ is the electron plasma 
frequency. Here the ions are assumed to be stationary. 

One can identify two distinct dispersion effects from (33): One long wavelength regime with
$ V_{Fe} \gg \hbar k/2 m_e, $ and the other short wavelength regime with $ V_{Fe}\leq \hbar k/2 m_e$.
These two regimes are separated by the critical wavenumber
\begin{equation}
k_{crit}=\frac{2\pi}{\lambda_{crit}} \approx \frac{\pi\hbar}{m_e V_{Fe}}\sim n_e^{-1/3}.
\end{equation}
It should be mentioned here that the quantum dispersion effects associated with the EPOs have recently
been observed in a compressed plasma \citep{Glenzer07,Neumayer10,Froula11}. In compressed plasma experiments,
powerful x-ray sources are employed for accessing narrow bandwidth electron plasma wave spectral
lines via collective Thomson scattering in which powerful light scatters off electron density fluctuations.
We note that the dispersion relation for EPOs in the finite electron temperature limit is given by \citep{Thiele08}
\begin{equation}
\omega^2 = \omega_{pe}^2+ 3 k^2 V_{Te}^2 (1+0.088 n_e \Lambda_e^3) + \frac{\hbar^2 k^4}{4 m_e^2},
\label{warm}
\end{equation}
where $V_{Te}=(k_B T_e/m_e)^{1/2}$ is the electron thermal speed and 
$\Lambda_e= \sqrt{2\pi} \hbar/{\sqrt{m_e k_B T_{e}}}$ the thermal (De Broglie) wavelength.

As mentioned in the Introduction, in the past many authors derived the dielectric constant for the
high-frequency (in comparison with the ion plasma frequency) ES waves \citep{Silin52a,Silin52b,Bohm1,Bohm2,Landau81} 
and the refractive index for EM waves \citep{Burt62} by using a quantum kinetic theory based on the Wigner 
and Poisson-Maxwell equations in a quantum plasma. In the following, we briefly discuss the well known results 
for the ES \citep{Silin52a,Silin52b,Bohm1,Bohm2} and EM \citep{Burt62} waves in an unmagnetized quantum plasma.

The dielectric constant for ES waves in a plasma with completely degenerate electrons reads
\citep{Landau81}
\begin{equation}
D_e(\omega, {\bf k}) = 1 + \frac{3\omega_{pe}^2}{2k^2 V_{Fe}^2}
\left[1- g(\omega_+) + g (\omega_-)\right],
\end{equation}
where $\omega_\pm =\omega \pm \hbar k^2/2m_e$, and
\begin{equation}
g(\omega_\pm ) =\frac{m_e (\omega_\pm^2 -k^2 V_{Fe}^2)}{2\hbar k V_{Fe}}
{\rm log} \left( \frac{\omega_\pm + kV_{Fe}}{\omega_\pm - kV_{Fe}} \right).
\end{equation}
Assuming that the phase velocity ($\omega/k)$ of the ES wave is much larger than $V_{Fe}$, we obtain
by setting $D_e(\omega, {\bf k}) =0$ the frequency of the EPOs, given by (36).
On the other hand, in the semi-classical limit, viz. $\hbar |{\bf k}| \ll p_{Fe} =\hbar(3\pi^2n_e)^{1/3}$,
we have \citep{Landau81} from Eq. (36)
\begin{equation}
D_e(\omega,{\bf k}) = 1+ \frac{3\omega_{pe}^2}{k^2 V_{Fe}^2}\left(1- \frac{\omega}{2 kV_{Fe}}
{\rm log} \left| \frac{\omega + k V_{Fe}}{\omega - k V_{Fe}} \right| \right),
\end{equation}
which in the short wavelength limit, viz. $k V_{Fe} \gg \omega_{pe}$, yields the so-called an electron 
thermal quasi-mode \citep{Silin52a,Silin52b,Silin61}
\begin{equation}
\omega = kV_{Fe}\left[1+ 2 \exp\left(-2k^2 \lambda_{s}^2-2\right)\right],
\end{equation}
where $\lambda_{s} = \lambda_{Fe}/\sqrt{3}$ is the Thomas-Fermi screening length.

Furthermore, when $\omega =0$, the expression (38) as a function of $k$ has a Kohn singularity
at $\hbar k = 2 p_{Fe} \equiv$ the diameter of the Fermi sphere. Here we have
\begin{equation}
D_e (0,{\bf k}) = 1+ \frac{e^2}{2\pi \hbar E_F}\left[1- \xi {\rm log}(1/|\xi|)\right],
\end{equation}
where $\xi=(\hbar k -2p_{Fe})/2p_{Fe}$ and $|\xi| \ll 1$. In a quantum plasma, with $D (0, {\bf k})$
given by (38), the potential distribution $\varphi (r)$ around a stationary test charge $q_t$ is
\begin{equation}
\varphi(r) = \frac{4 \pi q_t}{(2\pi)^3} \int \frac{\exp(i {\bf k}\cdot {\bf r}) d^3 k} { k^2 D_e (0,{\bf k})},
\end{equation}
which gives \citep{Else10}
\begin{equation}
\varphi(r) \approx q_t \frac{12 \lambda_{Fe}^2 \eta^4}{(2+3 \eta^2)^2} \frac{\cos(2k_F r)}{r^3},
\end{equation}
where $\eta =\hbar \omega_{pe}/4 k_B T_{Fe}$ and $k_F =p_{Fe}/\hbar$. We note that (42),
which is proportional to $r^{-3}\cos(2k_F r)$, considerably differs from the Debye-H\"uckel
shielding potential that is proportional to $r^{-1}\exp(-r/\lambda_{De})$ in a classical
plasma with the Maxwell-Boltzmann electron distribution function. Here $\lambda_{De}$ is
the electron Debye radius. We further note that the shielding of a moving test charge in an
unmagnetized quantum plasma has been investigated by \citet{Else10} both analytically and
numerically.

\subsection{Ion Plasma Oscillations (IPOs)}

We now focus our attention on the effect of the dynamics of non-relativistic and non-degenerate ions
in an unmagnetized quantum plasma. The dynamics of strongly coupled ions is governed by the ion hydrodynamic 
equations composed of Poisson's equation (28), and the continuity and momentum equations. The latter are
\begin{equation}
\frac{\partial n_i}{\partial t} + \nabla \cdot (n_i {\bf u}_i) =0,
\end{equation}
and
\begin{equation}
\begin{split}
&\left(1+ \tau_m \frac{\partial}{\partial t}\right) \bigg[\left(\frac{\partial}{\partial t} 
+ {\bf u}_i \cdot \nabla\right) {\bf u}_i
+ \frac{Z_i e}{m_i} \nabla \phi - \frac{\gamma_i k_B T_i}{m_i n_i}\nabla n_i \bigg]
  \\
&- \frac{\eta}{\rho_i} \nabla^2  {\bf u}_i - \frac{\left(\xi + \frac{\eta}{3}\right)}{\rho_i}
 \nabla (\nabla \cdot {\bf u}_i) =0,
\end{split}
\end{equation}
where $n_i$ is the ion number density, ${\bf u}_i$ the ion fluid velocity, $m_i$ the ion  mass, $\rho_i =n_i m_i$
the ion mass density, $\gamma_i$ the adiabatic index for the ion fluid, $\tau_m$ the viscoelastic relaxation time 
for ions, $\eta$ and $\xi$ the bulk ion viscosities. The viscoelastic equation (44) for strongly systems has been 
successfully used \citep{Ichimaru86,Kaw98} for investigating collective processes in classical plasmas with 
non-degenerate plasma particles.  

The ions are coupled with degenerate electrons by the space charge electric field ${\bf E} = -\nabla \phi$.
For low-phase velocity (in comparison with the electron Fermi speed) ES waves, we can neglect the inertia of 
the electrons to obtain
\begin{equation}
n_e \nabla \phi -  \frac{9}{5} \frac{\hbar^2 }{m_e} \nabla n_{e}^{5/3}
+ \frac{\hbar^2 n_e}{2m_e}\nabla \left(\frac{\nabla^2 n_e}{\sqrt{n_e}}\right) =0,
\end{equation}
for a quantum plasma with weakly relativistic degenerate electrons, while for a quantum plasma with
ultra-relativistic degenerate electrons, we have
\begin{equation}
 n_e \nabla \phi - \frac{3}{4} \hbar c \nabla n_e^{4/3} =0.
\end{equation}

Due to the ion inertia, one has new dielectric constants for the low-frequency (in comparison with the 
electron plasma frequency) ES waves \citep{Pines63,Pines89,Eliasson08a,Shukla08b,Mushtaq09}. In a quantum 
plasma with non-relativistic degenerate electrons with $\omega^2 \ll k^ 2 V_{Fe}^2 + \hbar^2 k^4/4m_e^2$,
we can linearize (28), (43), (44), and (45), Fourier transform them, and combine the resultant equations
to obtain
\begin{equation}
D_i (\omega, {\bf k}) = 1+ \frac{3 \omega_{pe}^2}{k^2 V_{Fe}^2 + \hbar^2 k^4/4m_e^2}
- \frac{\omega_{pi}^2} {\Omega_i^2},
\end{equation}
where $\omega_{pi} = (4\pi n_0 Z_i^2 e^2/m_i)^{1/2}$ is the ion plasma frequency, and $\Omega_i^2 =
\omega^2 - \gamma_i k^2 V_{Ti}^2 + i \omega k^2 \eta_*/(1-i \omega \tau_m))$, with $V_{Ti} =(k_B T_i/m_i)^{1/2}$ 
and $\eta_* = (\xi +4\eta/3)/m_i n_0$.  On the other hand, in a quantum plasma with ultra-relativistic 
degenerate electrons, we have from (28, (43), (44), and (46)
\begin{equation}
D_i(\omega, {\bf k}) = 1+ \frac{\omega_{pe}^2}{k^2 C_\hbar^2} - \frac{\omega_{pi}^2}{\Omega^2},
\end{equation}
where we have denoted $C_\hbar^2 = c^2 \lambda_C n_0^{1/3}$, and $\lambda_C =\hbar/m_e c$ is
the Compton length. By setting $D_i(\omega, {\bf k}) =0$, we obtain the frequencies of the IPOs.
For the case with non-relativistic degenerate electrons we have   
\begin{equation}
\omega^2  = \gamma_i k^2 V_{Ti}  + \frac{k^2 \eta_*}{\tau_m} 
+ \frac{\omega_{pi}^2 k^2 \lambda_{T\hbar}^2}{(1+ k^2 \lambda_{T\hbar}^2)},
\end{equation}
while for the case with ultra-relativistic degenerate electrons the result is 
\begin{equation}
\omega^2  = \gamma_i k^2 V_{Ti}^2 + \frac{k^2\eta_*}{\tau_m}  
+ \frac{\omega_{pi^2} k^2 \lambda_\hbar^2}{(1+ k^2 \lambda_\hbar^2)},
\end{equation}
where we have assumed $\omega \tau_m \ll 1$ and denoted
$\lambda_{T\hbar}  = \left[\lambda_s^2  + \hbar^2k^2 /4m_e^2 \omega_{pe}^2\right]^{1/2}$,
and $\lambda_\hbar =C_\hbar/\omega_{pe}$. The domain of validity of the hydrodynamic description
for the ions in the context of ion oscillations in a weakly relativistic dense plasma has also been 
recently discussed \citet{Mitten11}.

\citet{Melrose09} and \citet{Mushtaq09} have presented Landau damping rates for both electron and
ion plasma waves in an unmagnetized dense quantum plasma. The imaginary parts of the dielectric constants
can be used to calculate the structural form factor \citep{Ichimaru82} in quantum a plasma with degenerate 
electrons.

\citet{Shukla08b} used the dielectric constant (47) without the quantum statistical pressure term
(viz. the $V_{Fe}^2$-term) to investigate the screening and wake potentials around a test
charge in an electron-ion quantum plasma. They found a new screening potential \citep{Shukla08b}
\begin{equation}
\phi_{se} = \frac{q_t}{r}\exp(-k_q r) \cos(k_q r),
\end{equation}
and the wake potential
\begin{equation}
\phi_w = -\frac{q_t}{|z-u_0 t|}\cos\left[\frac{\omega_{pi}}{u_0}(z-u_0t)\right],
\end{equation}
where $k_q = \sqrt{2}/\sqrt{\hbar/m_e \omega_{pe}}$ is the quantum wave number, and
$r =[x^2 + y ^2+ (z-u_0t)^2]^{1/2}$  the distance from the test charge moving with the
speed $u_0$ along the $z$ axis in a Cartesian co-ordinate system. The wake potential (52) behind
a test charge arises due to collective interactions between a test charge and the ion oscillation
with the frequency $\omega_k \approx \omega_{pi}k_\perp /(k_\perp^2 + k_q^2)^{1/2}$,
with $k_z \ll k_q, k_\perp =(k_x^2 + k_y^2)^{1/2}$. We note that the Shukla-Eliasson (SE)
exponential cosine-screened Coulomb potential $\phi_{se}$ has a minimum of $\phi_{se} \approx -0.02 q_t k_q$
at $r\approx 3 k_q^{-1}$, similar to the Lennard-Jones potential for atoms. The SE screening potential
$\phi_{se}$, which is independent of the test charge speed $u_0$, is different from the Yukawa screening potential
$(q_t/r)\exp(-r/\lambda_s)$ that is valid in the limit $V_{Fe}\gg \hbar k /2 m_e$.  Recently, several
authors \citep{Ghoshal09a,Ghoshal09b,Xia10} have used the SE potential to study doubly
excited resonance states of Helium and hydrogen atoms embedded in a quantum plasma \citep{Ghoshal09a,Ghoshal09b},
and lattice waves in 2D hexagonal quantum plasma crystals \citep{Xia10}.

Furthermore, by using $D_i$ from (48), one can deduce potential distributions around a moving test
charge in a quantum plasma with ultra-relativistic electrons. We have
\begin{equation}
\phi (r,z) = \frac{q_t}{r}\exp \left(- \frac{r}{\Lambda_C}\right)
+ \frac{q_t}{|z-u_0t|}\cos\left[\frac{(z-u_0t)}{L_c}\right],
\end{equation}
where $\Lambda_C =C_\hbar/\omega_{pe}$ and $L_c = \lambda_c \left(M^2 -1 \right)^{1/2} > 0$, with
$M=u_0/C_\hbar$.

\subsection{High-Frequency EM Waves}

Finally, we turn our attention to the high-frequency (HF) EM waves in an unmagnetized quantum
plasma. Noting that HF-EM waves in the latter do not give rise to any density perturbations,
we have the EM wave frequency
\begin{equation}
\omega = (k^2 c^2 + \omega_{pe}^2)^{1/2}.
\end{equation}
However, consideration of the electron spin current and electron exchange potential contributions
in a quantum plasma gives rise to additional contributions to the refractive index $N$.
We have \citep{Burt62}
\begin{equation}
\frac{k^2 c^2}{\omega^2 } = N \approx 1- \frac{\omega_{pe}^2}{\omega^2 }
-\frac{\omega_{pe}^2 \hbar^2 k^2}{m_e^2 \omega^4}\left(\frac{1}{5}K_F^2 + \frac{1}{4}k^2\right),
\end{equation}
which includes the electron spin correction, and is valid at zero temperature. Here
$\hbar K_F =(2m_e E_{Fe})^{1/2}$ is the momentum of degenerate electrons at
the Fermi surface, the $(1/5) K_F^2$ term is related to the leading quantum
term from the ordinary transverse current, and the $k^2/4$ term arises from the electron
spin interactions. On the other hand, the EM wave dispersion relation, which
accounts for the electron exchange potential and discards the spin correction, reads \citep{Burt62}
\begin{equation}
\frac{k^2 c^2}{\omega^2} = N \approx 1- \frac{\omega_{pe}^2}{\omega^2 }
-\frac{\omega_{pe}^2 \hbar^2 k^2 K_F^2} {5m_e^2\omega^4} + \frac{3\omega_{pe}^2 k^2}{40 \omega^4K_F^2}.
\end{equation}

\section{Quantum Dark Solitons and Vortices}

Let us now discuss nonlinear properties and dynamics of 1D quantum dark solitons (characterized by 
the local electron density depletion associated with a positive potential) and 2D azimuthally symmetric 
electron vortices in an unmagnetized quantum plasma \citep{ShuklaEliasson06} with immobile ions. 
The assumption of stationary ions is justified because we are looking for the nonlinear phenomena on 
a timescale much shorter than the ion plasma period.

We use the normalized NLS-Poisson equations \citep{Shukla06,ShuklaEliasson06}
\begin{equation}
i \frac{\partial \Psi}{\partial t} + {\cal A} \nabla^2 \Psi + \varphi \Psi - |\Psi|^{4/D} \Psi =0,
\label{Eq1}
\end{equation}
and
\begin{equation}
\nabla^2\varphi = |\Psi|^2 -1,
\label{Eq2}
\end{equation}
where the time and space variables are in units of $\hbar/k_B T_{Fe}$ and the electron Fermi-Thomas
screening length $\lambda_{TF}$, respectively. Furthermore, we have denoted $\Psi=\psi/\sqrt{n_0}$,
$\varphi=e\phi/k_B T_{Fe}$, and ${\cal A}=2\pi n_0^{1/3}e^2/k_B T_{Fe}$. The system (\ref{Eq1}) and
(\ref{Eq2}) is supplemented by
\begin{equation}
\frac{\partial {\bf E}_\varphi}{\partial t}= i {\cal A}\left(\Psi\nabla\Psi^*-\Psi^*\nabla\Psi\right),
\label{Eq3}
\end{equation}
where ${\bf E}_\varphi=-\nabla \varphi$. Equations (\ref{Eq1})--(\ref{Eq3}) have the
following conserved integrals \citep{ShuklaEliasson06,Shaikh07}: the number of electrons
\begin{equation}
  N=\int |\Psi|\, d^3 x,
\end{equation}
the electron momentum
\begin{equation}
{\bf P}=-i\int \Psi^*\nabla\Psi\, d^3 x,
\end{equation}
the electron angular momentum
\begin{equation}
{\bf L}=-i\int \Psi^*{\bf r}\times\nabla\Psi\, d^3 x,
\end{equation}
and the total energy
\begin{equation}
{\cal E}=\int \left[- {\cal A} \Psi^* \nabla^2\Psi + \frac{|\nabla\varphi|^2}{2}
+ \frac{D}{(2 +D)} |\Psi|^{(2+4/D)} \right] \,d^3 x.
\end{equation}

\subsection{Quantum Electron Cavity}

For quasi-stationary, 1D nonlinear structures moving with a constant speed $v_0$, one can
find solitary wave solutions of Eqs. (57) and (58) by introducing the ansatz $\Psi=W(\xi)\exp(iK_s x-i\Omega_s t)$,
where $W$ is a complex-valued function of the argument $\xi=x-v_0t$, and $K_s$ and $\Omega_s$ are a
constant wavenumber and frequency shift, respectively. By the choice $K_s= v_0/2{\cal A}$, the
coupled system of equations (57) and (58) can then be written as
\begin{equation}
 \frac{d^2 W}{d\xi^2}+\lambda W+\frac{\varphi W}{{\cal A}}-\frac{|W|^4 W}{{\cal A}}=0,
\label{Eq4}
\end{equation}
and
\begin{equation}
\frac{d^2\varphi}{d\xi^2}=|W|^2-1,
\label{Eq5}
\end{equation}
where $\lambda=(\Omega_s/ {\cal A})-v_0^2/4{\cal A}^2$ is an eigenvalue of the system.
From the boundary conditions $|W|=1$ and $\varphi=0$ at $|\xi|=\infty$,
we determine $\lambda=1/{\cal A}$ and $\Omega_s=1+v_0^2/4 {\cal A}$. The system of
Eqs. (\ref{Eq4}) and (\ref{Eq5}) admits a first integral in the form
\begin{equation}
  \begin{split}
  H_h = &{\cal A}\left|\frac{dW}{d\xi}\right|^2 -\frac{1}{2}\left(\frac{d\varphi}{d\xi}\right)^2 +|W|^2
  \\
  &-\frac{|W|^6}{3} +\varphi |W|^2-\varphi-\frac{2}{3}=0,
  \end{split}
  \label{Eq6}
\end{equation}
where the boundary conditions $|W|=1$ and $\varphi=0$ at $|\xi|=\infty$ have been employed.

\begin{figure}[htb]
\centering
\includegraphics[width=8cm]{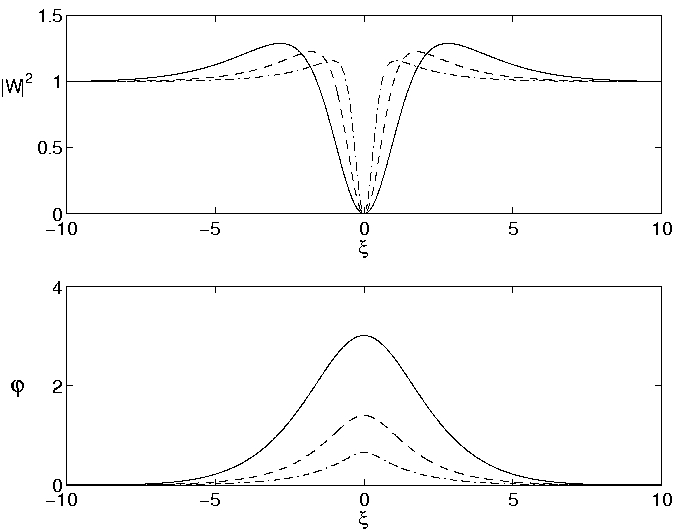}
\caption{The electron density $|W|^2$ (the upper panel) and ES potential $\varphi$ (the lower panel)
associated with a dark soliton supported by the system of equations (\ref{Eq4}) and (\ref{Eq5}), for ${\cal A}=5$
(solid lines), ${\cal A}=1$ (dashed lines), and ${\cal A}=0.2$ (dash-dotted line). After \citet{ShuklaEliasson06}.}
\label{Fig1}
\end{figure}

\begin{figure}[htb]
\centering
\includegraphics[width=8cm]{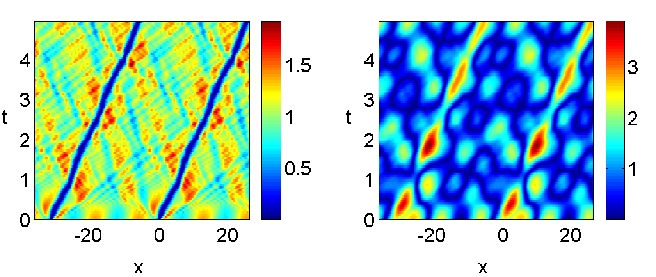}
\caption{The time-development of the electron density $|\Psi|^2$ (left-hand panel) and ES
potential $\varphi$ (the right-hand panel), obtained from a simulation of the system of
equations (\ref{Eq1}) and (\ref{Eq2}).  The initial condition is $\Psi=0.18+\tanh[20\sin(x/10)]\exp(i K_s x)$,
with $K_s= v_0/2{\cal A}$, ${\cal A}=5$ and $v_0=5$.  After \citet{ShuklaEliasson06}.}
\label{Fig2}
\end{figure}

Figure \ref{Fig1} shows profiles of $|W|^2$ and $\varphi$ obtained numerically from (\ref{Eq4}) and (\ref{Eq5})
for a few values of ${\cal A}$, where $W$ was set to $-1$ on the left boundary and to $+1$ on the right boundary,
i.e. the phase shift is 180 degrees between the two boundaries. The solutions are in the form of
dark solitons, with a localized depletion of the electron density $N_e=|W|^2$, associated with
a localized positive potential. Larger values of the quantum coupling parameter
$A$ give rise to larger-amplitude and wider dark solitons. The solitons localized ``shoulders''
on both sides of the density depletion.

A numerical solution of the time-dependent system of Eqs. (\ref{Eq1}) and (\ref{Eq2}) is displayed
in Fig. \ref{Fig2}, with initial conditions close (but not equal) to the ones in Fig. \ref{Fig1}.
Two very clear and long-lived dark solitons are visible, associated with a positive
potential of $\varphi\approx 3$, in agreement with the quasi-stationary solution of
Fig. \ref{Fig1} for ${\cal A} = 5$. In addition there are oscillations and wave turbulence in the
time-dependent solution presented in Fig. \ref{Fig2}.  Hence, the dark solitons seem to be
robust structures that can withstand perturbations and turbulence during a considerable time.

\begin{figure}[htb]
\centering
\includegraphics[width=8cm]{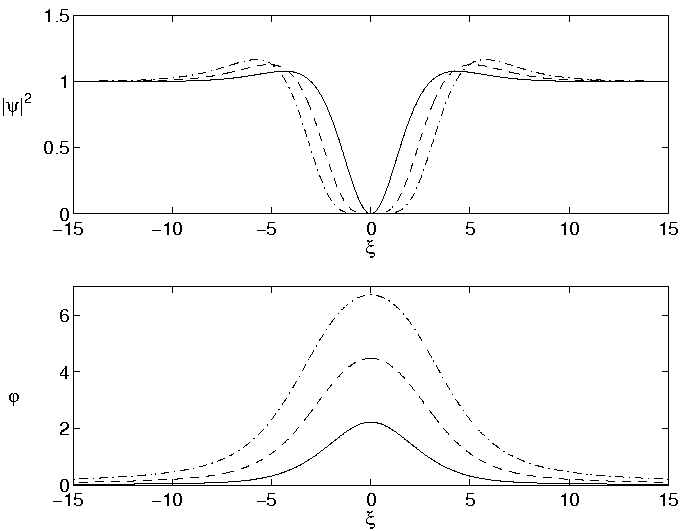}
\caption{The electron density $|\Psi|^2$ (upper panel) and ES potential $\varphi$ (lower panel)
associated with 2D electron vortices supported by the system (\ref{Eq7}) and (\ref{Eq8}), for the
charge states $s=1$ (solid lines), $s=2$ (dashed lines) and $s=3$ (dash-dotted lines), with ${\cal A} = 5$
in all cases. After \citet{ShuklaEliasson06}.}
\label{Fig3}
\end{figure}

\begin{figure}[htb]
\centering
\includegraphics[width=8cm]{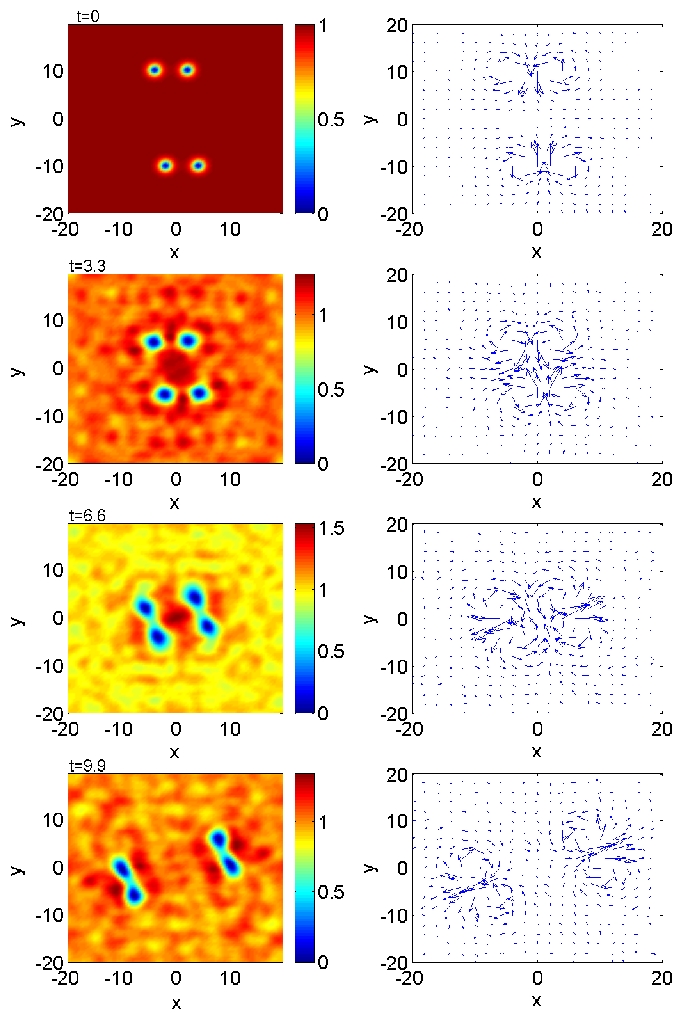}
\caption{The electron density $|\Psi|^2$ (left panel) and an arrow plot of the electron current
$i\left(\Psi\nabla\Psi^*-\Psi^*\nabla\Psi\right)$ (right panel) associated with singly charged
($|s|=1$) 2D electron vortices, obtained from a simulation of the time-dependent system of equations
(\ref{Eq1}) and (\ref{Eq2}), at times $t=0$, $t=3.3$, $t=6.6$ and $t=9.9$ (upper to lower panels), with
${\cal A} = 5$. The singly charged vortices form pairs and keep their identities.
After \citet{ShuklaEliasson06}.}
\label{Fig4}
\end{figure}

\begin{figure}[htb]
\centering
\includegraphics[width=8cm]{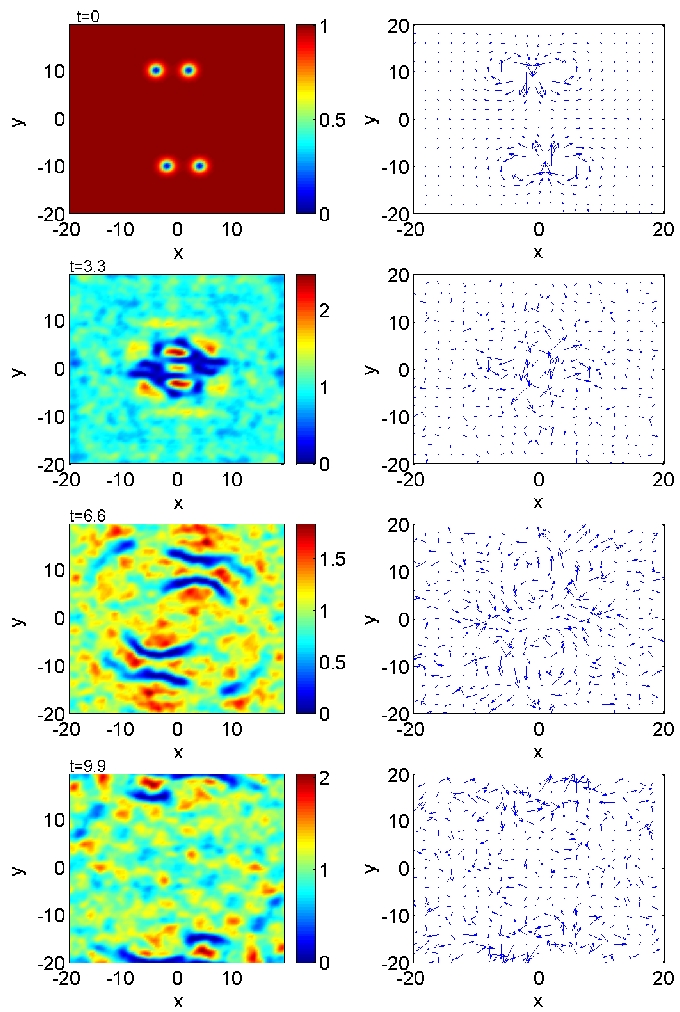}
\caption{The electron density $|\Psi|^2$ (left panel) and an arrow plot of the electron current
$i\left(\Psi\nabla\Psi^*-\Psi^*\nabla\Psi\right)$ (right panel) associated with double charged
($|s|=2$) 2D electron vortices, obtained from a simulation of the time-dependent system of Eqs.
(\ref{Eq1}) and (\ref{Eq2}), at times $t=0$, $t=3.3$, $t=6.6$ and $t=9.9$ (upper to lower panels),
with ${\cal A} = 5$. The doubly charged vortices dissolve into nonlinear structures and wave turbulence.
After \citet{ShuklaEliasson06}.}
\label{Fig5}
\end{figure}

\subsection{Quantum Electron Vortices}

For two-dimensional ($D=2$) EPOs in quantum plasmas, one can look for quantum vortex structures
of the form $\Psi=\psi(r)\exp(is\theta-i\Omega_v t)$, where $r$ and $\theta$ are the polar
coordinates defined via $x=r\cos(\theta)$ and $y=r\sin(\theta)$, $\Omega_v$ is a constant frequency
shift, and $s=0, \, \pm 1,\, \pm 2,\ldots$ for different excited states (charge states).
With this ansatz, Eqs. (\ref{Eq1}) and (\ref{Eq2}) can be written as, respectively,
\begin{eqnarray}
&&\!\!\!\!\!\!\!\!\!\!\!\!\!\!\!\!\!\!\!\!\!\!\!\!\!\!\!\left[\Omega_v
+{\cal A}\left(\frac{d^2}{dr^2}+\frac{1}{r}\frac{d}{dr}-\frac{s^2}{r^2}\right)
+\varphi-|\psi|^2\right]\psi=0,
\label{Eq7}
\end{eqnarray}
and
\begin{equation}
\left(\frac{d^2}{dr^2}+\frac{1}{r}\frac{d}{dr}\right)\varphi=|\psi|^2-1,
\label{Eq8}
\end{equation}
where the boundary conditions $\psi=1$ and $\varphi=d\psi/dr=0$ at $r=\infty$ determine the
constant frequency $\Omega_v = 1$. Different signs of the charge state $s$ describe different
rotation directions of the quantum vortex.  For $s\neq 0$, one must have $\psi=0$ at $r=0$,
and from symmetry considerations one has $d\varphi/dr=0$ at $r=0$. Figure \ref{Fig3} depicts
numerical solutions of Eqs. (\ref{Eq7}) and (\ref{Eq8}) for different values of $s$ and
for $A=5$. Here a quantum vortex is characterized by a complete depletion of the electron
density at the core of the vortex, and is associated with a positive ES potential.

A time-dependent solution of Eqs. (\ref{Eq1}) and (\ref{Eq2}) in two-space dimensions for singly 
charged ($s=\pm 1$) electron vortices is shown in Fig. \ref{Fig4}, where, in the initial
condition, four vortex-like structures were placed at some distance from each other.
The initial conditions were such that the vortices are organized in two vortex pairs,
with $s_1=+1$, $s_2=-1$, $s_3=-1$, and $s_4=+1$, seen in the upper panels of Fig. \ref{Fig4}.
The vortices in the pairs have opposite polarity on the electron fluid rotation, as seen in
the upper right panel of Fig. \ref{Fig4}. Interestingly, the ``partners'' in the vortex pairs
attract each other and propagate together with a constant velocity, and in the collision and interaction
of the vortex pairs (see the second and third pairs of panels in Fig. \ref{Fig4}), the vortices
keep their identities and change partners, resulting into two new vortex pairs which propagate
obliquely to the original propagation direction. On the other hand, as shown in Fig. \ref{Fig5},
vortices that are multiply charged ($|s_j|>1$) are unstable.  Here the system of Eqs. (\ref{Eq1})
and (\ref{Eq2}) was again solved numerically with the same initial condition as the one in
Fig. \ref{Fig4}, but with doubly charged vortices $s_1=+2$, $s_2=-2$, $s_3=-2$, and $s_4=+2$.
The second row of panels in Fig. \ref{Fig5} reveals that the vortex pairs keep their identities
for some time, while a quasi 1D density cavity is formed between the two vortex pairs.
At a later stage, the four vortices dissolve into complicated nonlinear structures and wave turbulence.
Hence, the nonlinear dynamics is very different between singly and multiply charged solitons, where only
singly charged vortices are long-lived and keep their identities.

\section{Quantum Electron Fluid Turbulence\label{sec:turbulence}}

The statistical properties of quantum electron fluid turbulence and its associated electron
transport properties at nanoscales in a quantum plasma have been investigated in both 2D and
3D by using the coupled NLS and Poisson equations \citep{Shaikh07,Shaikh08}. It has been found
that nonlinear couplings between the EPOs of different scale sizes give rise to
small-scale electron density structures, while the ES potential cascades
towards large-scales. The total energy associated with the quantum electron plasma
wave turbulence processes a non-universal spectrum that depends on the
quantum electron coupling parameter.

To investigate 3D quantum electron plasma wave turbulence, we use the NLS-Poisson
equations \citep{Manfredi01,Shukla06,ShuklaEliasson06,Shaikh08}
\begin{equation}
i \sqrt{2H} \frac{\partial \Psi}{\partial t}+ H \nabla^2\Psi
+ \varphi \Psi - |\Psi|^{4/3}\Psi = 0,
\label{Eq1D}
\end{equation}
and
\begin{equation}
\nabla^2\varphi = |\Psi|^2-1,
\label{Eq2D}
\end{equation}
were used, which govern the dynamics of nonlinearly interacting EPOs of different wavelengths. In
Eqs. (\ref{Eq1D}) and (\ref{Eq2D}) the wavefunction is normalized by $\sqrt{n_0}$, the ES
potential by $k_B T_{Fe}/e$, the time $t$ by the electron plasma period $\omega_{pe}^{-1}$,
and the space ${\bf r}$ by the electron Fermi-Thomas screening length $\lambda_{Fe} =V_{Fe}/\omega_{pe}$.
Here $\sqrt{H} = \hbar \omega_{pe}/\sqrt{2} k_B T_{Fe}$ was introduced.

The nonlinear wave-wave coupling studies have been performed to investigate the
multi-scale evolution of a decaying 3D electron plasma wave turbulence, which is described by
Eqs. (\ref{Eq1D}) and (\ref{Eq2D}). All fluctuations are initialized isotropically
(no mean fields are assumed) with random phases and amplitudes in Fourier space,
and are evolved in time by the integration of Eqs. (\ref{Eq1D}) and (\ref{Eq2D}) numerically.
The initial isotropic turbulent spectrum was initially chosen close to $k^{-2}$, with
random phases in all three directions. The choice of such (or even a flatter than $k^{-2}$)
spectrum treats the turbulent fluctuations on an equal footing and avoids any influence
on the dynamical evolution that may be due to the initial spectral non-symmetry.

\begin{figure}[htb]
\centering
\includegraphics[width=8cm]{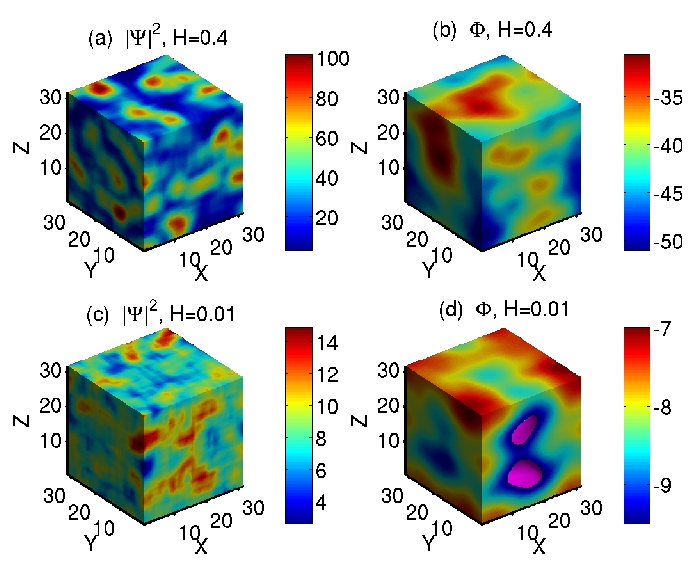}
\caption{Small-scale fluctuations in the electron density resulting from steady turbulence
simulations, for $H=0.4$ (top panels) and $H=0.01$ (bottom panels).
Forward cascades are responsible for the generation of small-scale fluctuations seen in panels
(a) and (c).  Large scale structures are present in the ES potential, seen in panels
(b) and (d), essentially resulting from an inverse cascade. After \citet{Shaikh08}.}
\label{Fig1D}
\end{figure}

The properties of 3D electron plasma wave turbulence, composed of nonlinearly interacting EPOs,
were studied for two specific physical systems, corresponding to dense plasmas in the next generation of
laser-based plasma compression (LBPC) schemes \citep{Malkin07}, and in superdense astrophysical
objects \citep{Chabrier02,Chabrier06,Chabrier09,Lai01,Harding06} (e.g. white dwarfs).  It is expected that
in LBPC schemes, the electron number density may reach $10^{27}$ cm$^{-3}$ and beyond.
Hence, we have $\omega_{pe} =1.76 \times 10^{18}$ s$^{-1}$, $T_{Fe} = 1.7 \times 10^{-9}$
erg, $\hbar \omega_{pe} = 1.7 \times 10^{-9}$ erg, and $H = 1$, and the electron Fermi-Thomas screening length
$\lambda_{Fe} = 0.1$~\AA.  On the other hand, in the core of white dwarf stars, we typically
have $n_0 \sim 10^{30}$ cm$^{-3}$, yielding $\omega_{pe} =5.64 \times 10^{19}$ s$^{-1}$,
$T_{Fe} = 1.7 \times 10^{-7}$ erg (0.1 MeV), $\hbar \omega_{pe} = 5.64 \times 10^{-8}$ erg,
$H  \approx 0.3$, and $\lambda_{Fe} = 0.025$~\AA.  The numerical solutions of Eqs. (\ref{Eq1D})
and (\ref{Eq2D}) for $H=0.4$ and $H=0.01$ (corresponding to $n_0 =10^{27}$ cm$^{-3}$ and
$n_0 =10^{30}$ cm$^{-3}$, respectively) are displayed in Fig. \ref{Fig1D}, which shows the 
electron number density and ES potential distributions in the $(x,y,z)$-cube.

\begin{figure}[htb]
\centering
\includegraphics[width=8cm]{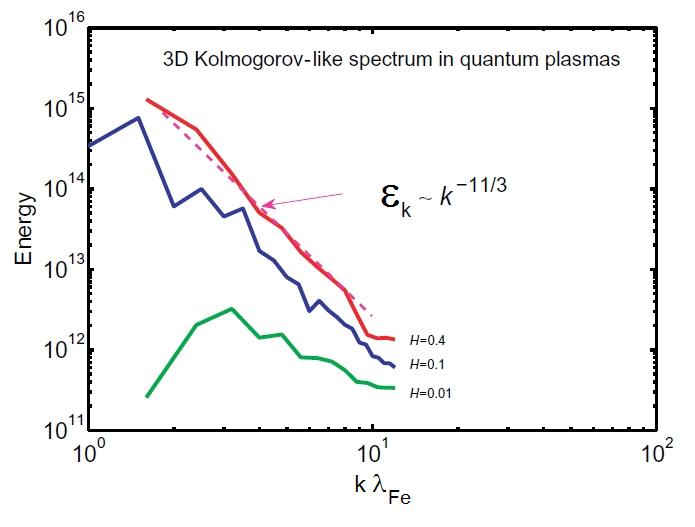}
\caption{ Energy ${\cal E}_k$ per vector wavenumber of 3D EPOs in the forward cascade regime.
A Kolmogorov-like spectrum ${\cal E}_k\sim k^{-11/3}$ is observed for $H=0.4$. The spectral
index changes as a function of $H$.  After \citet{Shaikh08}.}
\label{Fig2D}
\end{figure}

Figure \ref{Fig1D} reveals that the electron density distribution has a tendency to generate smaller
length-scale structures, while the ES potential cascades towards larger scales. The co-existence of the
small and larger scale structures in turbulence is a ubiquitous feature of various 3D turbulence systems.
For example, in 3D hydrodynamic turbulence, the incompressible fluid admits two invariants, namely the
energy and the mean squared vorticity.  The two invariants, under the action of an external forcing,
cascade simultaneously in turbulence, thereby leading to a dual cascade phenomena. In these processes,
the energy cascades towards longer length-scales, while the fluid vorticity transfers spectral power
towards shorter length-scales.  Usually, a dual cascade is observed in a driven turbulence simulation,
in which certain modes are excited externally through random turbulent forces in spectral space.
The randomly excited Fourier modes transfer the spectral energy by conserving the constants of motion
in $k$-space. On the other hand, in freely decaying turbulence, the energy contained in the large-scale
eddies is transferred to the smaller scales, leading to a statistically stationary inertial regime associated
with the forward cascades of one of the invariants. Decaying turbulence often leads to the formation of
coherent structures as turbulence relaxes, thus making the nonlinear interactions rather inefficient when
they are saturated.  The power spectrum exhibits an interesting feature in the 3D electron plasma
system discussed here, unlike the 3D hydrodynamic turbulence \citep{Kolmogorov41a,Kolmogorov41b,Lesieur90,Frisch95}.
Figure \ref{Fig2D} shows the energy ${\cal E}_k$ per vector wavenumber. For isotropic
3D turbulence, it is related to the energy $E_k$ per scalar wavenumber as $E_k=4\pi k^2 {\cal E}_k$;
see e.g. \citet{Knight90}.  For $H=0.4$, the spectrum per vector wavenumber is close to
${\cal E}_k\sim k^{-11/3}$ and hence yields the standard Kolmogorov power spectrum \citep{Kolmogorov41a,Kolmogorov41b}
$E_k\sim k^{-5/3}$. However, the spectrum is not universal but changes for different values of $H$.
For 2D quantum electron fluid turbulence \citep{Shaikh07} the spectral slope was more close to the
Iroshnikov-Kraichnan power law \citep{Iroshnikov63,Kraichnan65} $E_k \sim k^{-3/2}$. The origin
of the differences in the observed spectral indices resides with the nonlinear character of the
underlying plasma models, as nonlinear interactions in the 2D and 3D systems are governed typically
by different nonlinear forces. The latter modify the spectral evolution of turbulent cascades
to a significant degree.  Physically, the flatness (or deviation from the $k^{-5/3}$ law), results
from the short wavelength part of the EPOs spectrum which is controlled by the quantum electron
tunneling effect associated with the Bohm potential.  The peak in the energy spectrum can be
attributed to the higher turbulent power residing in the EPO potential, which eventually leads
to the generation of larger scale structures, as the total energy encompasses both the electrostatic
potential and electron density components.  In the dual cascade process, there is a delicate competition
between the EPO dispersions caused by the statistical pressure law (giving the $k^2 V_{Fe}^2$ term,
which dominates at longer scales) and the quantum Bohm force (giving the $\hbar^2k^4/4m_e^2$ term,
which dominates at shorter scales).

\begin{figure}[htb]
\centering
\includegraphics[width=8cm]{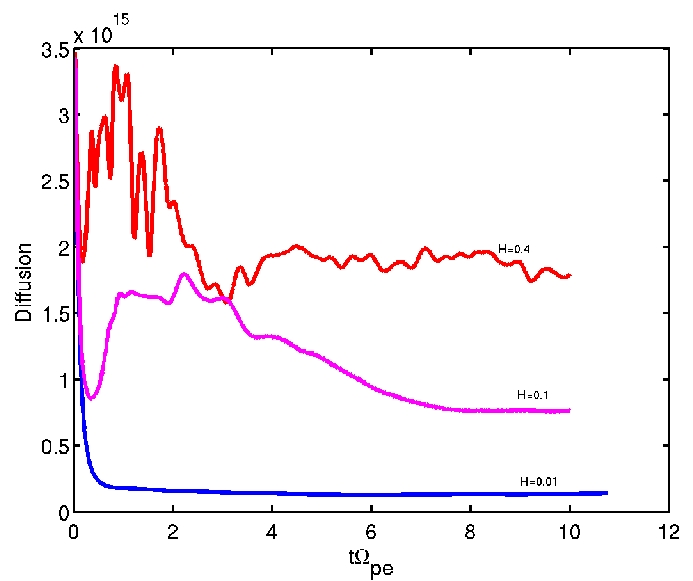}
\caption{Time evolution of the effective electron diffusion coefficient associated with the
large-scale ES potential and the small-scale electron density, for
$H=0.4$, $H=0.1$ and $H=0.01$. Smaller values of $H$ corresponds to a small effective
diffusion coefficient, which characterizes the presence of small-scale turbulent eddies
that suppress the electron transport.  After \citet{Shaikh08}.}
\label{Fig3D}
\end{figure}

The electron diffusion in the presence of small and large scale turbulent EPOs can be estimated in the 
following manner. An effective electron diffusion coefficient caused by the momentum transfer can be calculated 
from $D_{eff} = \int_0^\infty \langle {\bf P}({\bf r},t) \cdot {\bf P}({\bf r},t+ t^\prime) \rangle dt^\prime$,
where ${\bf P}$ is electron momentum and the angular bracket denotes spatial averages and the ensemble
averages are normalized to unit mass. The effective electron diffusion coefficient, $D_{eff}$, essentially
relates the diffusion processes associated with random translational motions of the electrons in
nonlinear plasmonic fields. To measure the turbulent electron transport that is associated with the turbulent 
structures, $D_{eff}$ is computed. It is observed that the effective electron diffusion is lower when the field 
perturbations are Gaussian. On the other hand, the electron diffusion increases rapidly with the eventual formation 
of large-scale structures, as shown in Fig. \ref{Fig3D}.  The electron diffusion due to large scale potential 
distributions in a quantum plasma dominates substantially, as depicted by the solid-curve in Fig. \ref{Fig3D}.
Furthermore, in the steady-state, nonlinearly coupled EPOs form stationary structures, and $D_{eff}$ eventually
saturates. Thus, remarkably an enhanced electron diffusion results primarily due to the emergence of
large-scale potential structures.

\section{Nonlinearly Coupled EM and ES Waves}

We turn our attention to nonlinear interactions between large amplitude EM and
ES waves in a  quantum plasma.  \citet{Shukla2006} considered nonlinear couplings
between large amplitude EM waves and finite amplitude electron and ion plasma waves,
and presented nonlinear dispersion relations that exhibit stimulated Raman scattering (SRS),
stimulated Brillouin scattering (SBS), and modulational instabilities.  The work of 
\citet{Shukla2006} has been further generalized by including thermal corrections
to the ES waves \citep{Stenflo09} and relativistic electron mass variations \citep{Shukla07}
caused by EM waves in an unmagnetized  quantum plasma.

\subsection{Stimulated Scattering Instabilities}

First, we present the governing equations for the HF-EM waves and
the EM wave driven modified EPOs and IPOs.
We have \citep{Stenflo09}
\begin{equation}
\left(\frac{\partial^2}{\partial t^2} -c^2 \nabla^2 + \omega_{pe}^2\right) {\bf A}+
\omega_{pe}^2 \frac{n_1}{n_0}{\bf A} \approx 0,
\end{equation}
for the HF-EM wave,
\begin{equation}
\left(\frac{\partial^2}{\partial t^2} + \omega_{pe}^2  -\frac{3}{5} V_{Fe}^2 \nabla^2
+ \frac{\hbar^2} {4 m_e^2} \nabla^4\right) \frac{n_1}{n_0} = \frac{e^2}{2m_e^2c^2} \nabla|{\bf A}|^2,
\end{equation}
for the HF-EM wave pressure driven EPOs, and
\begin{equation}
\left(\frac{\partial^2}{\partial t^2} -C_{TF}^2 \nabla^2 + \frac{\hbar^2}
{4 m_e m_i} \nabla^4\right) \frac{n_1}{n_0} = \frac{e^2}{2m_e m_i c^2} \nabla|{\bf A}|^2,
\end{equation}
for the EM wave pressure driven modified IPOs without the ion thermal, ion viscoelastic relaxation,
and ion viscosity effects. Here $C_{TF} =(k_B T_{Fe}/m_i)^{1/2}$ and $n_1 (\ll n_0)$ is a small perturbation 
in the electron number density.

Following the standard procedure of the parametric instabilities
\citep{Yu74,Shukla81,Sharma83,Murtaza84,Shukla06,ShuklaEliasson06},
we can Fourier analyze (71)-(73) and combine the resultant equations to obtain the nonlinear dispersion
relations
\begin{equation}
\omega^2 - \Omega_R^2 = - \frac{e^2\omega_{pe}^2k^2 |{\bf A}_0|^2}{2m_e^2 c^2} \left(\frac{1}{D_{+}}+\frac{1}{D_{-}}\right),
\end{equation}
and
\begin{equation}
\omega^2 - \Omega_B^2 =\frac{e^2\omega_{pe}^2k^2 |{\bf A}_0|^2}{2m_e m_i c^2}\left(\frac{1}{D_{+}}+\frac{1}{D_{-}}\right),
\end{equation}
for the driven EPOs and IPOs, respectively, which admit SRS, SBS, and modulational instabilities of 
the HF-EM pump (with the amplitude ${\bf A}_0$) in a quantum plasma.  Here, 
$D_{\pm} = \pm 2\omega_0(\omega-{\bf k}\cdot {\bf V}_g) - k^2 c^2$,
where ${\bf V}_g ={\bf k} c^2/2\omega_0$ is the group velocity of the HF-EM pump wave with the frequency
$\omega_0 =(k_0^2 c^2+\omega_{pe}^2)^{1/2}$, and
\begin{equation}
\Omega_R^2 = \omega_{pe}^2 +\frac{3}{5}k^2 V_{Fe}^2 +\frac{\hbar^2 k^4}{4m_e^2},
\end{equation}
and
\begin{equation}
\Omega_B^2 = k^2 C_{TF}^2 + \frac{\hbar^2 k^4}{4m_e m_i}.
\end{equation}
The growth rates of SRS and SBS instabilities \citep{Shukla2006} are, respectively,
\begin{equation}
\gamma_R =\frac{\omega_{pe}e K|{\bf A}_0|}{2\sqrt{2 \omega_0\Omega_R} m_ec},
\end{equation}
and
\begin{equation}
\gamma_B =\frac{\omega_{pe}e K|{\bf A}_0|}{2\sqrt{2 \omega_0\Omega_B m_e m_i} c}.
\end{equation}
The present results of SRS and SBS instabilities will help to identify the electrostatic spectral
lines that are enhanced by the large amplitude HF-EM pump wave in a quantum plasma.

\subsection{Nonlinearly Coupled Intense EM and EPOs}

Let us now consider nonlinear interactions between an arbitrary large amplitude circularly polarized
electromagnetic (CPEM) wave and nonlinear EPOs that are driven by the relativistic ponderomotive
force \citep{Shukla84,Shukla86} of the CPEM waves. Such an  interaction gives rise to an envelope
of the CPEM vector potential ${\bf A}_\perp=A_\perp (\hat{\bf x}+ i\hat{\bf y}) \exp(-i\omega_0 t+ i k_0 z)$,
which obeys the NLS equation \citep{Shukla07}
\begin{equation}
2i \epsilon \left(\frac{\partial}{\partial t} + U_g\frac{\partial}{\partial z}\right) A_\perp
+\frac{\partial^2A_\perp}{\partial z^2}-\left(\frac{|\psi|^2}{\gamma}-1\right)A_\perp=0,
\label{Eq9}
\end{equation}
where $\epsilon =\omega_0/\omega_{pe}$, and the normalized (by $\sqrt{n_0}$) electron wavefunction
$\psi$ and the normalized (by $m_0c^2/e$) scalar potential are governed by, respectively,
\begin{equation}
 i H_e \frac{\partial \psi}{\partial t} +\frac{H_e^2}{2}\frac{\partial^2\psi}{\partial z^2}
+(\phi-\gamma+1)\psi=0,
 \label{Eq10}
\end{equation}
and
\begin{equation}
\frac{\partial^2\phi}{\partial z^2}=|\psi|^2-1,
\label{Eq11}
\end{equation}
where  $m_0$ is the rest mass of the electrons, $U_g =k_0 c /2\omega_0$ $ H_e=\hbar\omega_{pe}/m_0 c^2$
is the ration between the plasmonic energy density to the rest electron energy, and
$\gamma=(1+|{A}_\perp|^2)^{1/2}$ is the relativistic gamma factor due to the electron quiver velocity
in the CPEM wave fields. The time and space variables are in units of the electron plasma period
($\omega_{pe}^{-1}$) and the electron skin depth $\lambda_e = c/\omega_{pe}$. The electron density
and $A_\perp$ are in units of $n_0$ and $m_0c^2/e$ \citet{Shukla07}. The nonlinear coupling between
intense CPEM waves and EPOs comes about due to the nonlinear current density, which is represented by
the term $|\psi|^2 {A}_\perp/\gamma$ in Eq. (\ref{Eq9}). In Eq. (\ref{Eq10}), $1-\gamma$
is the relativistic ponderomotive potential \citep{Shukla84,Shukla86}. The latter arises from
the averaging (over the CPEM wave period $2\pi/\omega_0$) of the relativistic advection and the
nonlinear Lorentz force involving the electron quiver velocity and the CPEM wave electric and
magnetic fields.

A relativistically strong EM wave in a classical electron-ion plasma is subject to SRS and
modulational instabilities \citep{McKinstrie92}. One can expect that these instabilities will
be modified at quantum scale by the dispersion effects caused by the tunneling of
electrons through the quantum Bohm potential. The growth rate of the relativistic parametric
instabilities in a dense quantum plasma in the presence of a relativistically strong
CPEM pump wave can be obtained in a standard manner \cite{Shukla86} by letting $\phi(z,t)=\phi_1(z,t)$,
${A}_\perp(z,t) = [{A}_0+ {A}_1(z,t)]\exp(-i\alpha_0 t)$ and $\psi(z,t)=[1+\psi_1(z,t)]\exp(-i\beta_0 t)$,
where ${A}_0$ is the large-amplitude CPEM pump and ${A}_1$ is the small-amplitude perturbation
of the CPEM wave amplitude due to the nonlinear coupling between the CPEM waves and EPOs,
i.e. $|{A}_1|\ll |{A}_0|$, and $\psi_1$ $(\ll 1)$ is the small-amplitude perturbation in the
electron wave function. The constants $\alpha_0$ and $\beta_0$ are constant frequency shifts,
determined from Eqs. (\ref{Eq9}) and (\ref{Eq10}) to be $\alpha_0=(1/\gamma_0-1)/(2\epsilon)$,
and $\beta_0=(1-\gamma_0)/H_e$, where $\gamma_0=(1+|{A}_0|^2)^{1/2}$. The first-order perturbations
in the electromagnetic vector potential and the electron wave function are expanded into their
respective sidebands as ${A}_1(z,t)={A}_+\exp(i K z-i\Omega t) +{A}_{-}\exp(-i K z+i\Omega t)$
and $\psi_1(z,t)=\psi_+\exp(i K z-i\Omega t)+\psi_{-}\exp(-iKz+i\Omega t)$, while the potential
is expanded as $\phi(z,t)=\widehat{\phi}\exp(iKz-i\Omega t)+\widehat{\phi}^{*} \exp(-iKz+i\Omega t)$,
where $\Omega$ and $K$ are the normalized frequency and the normalized wave number of the EPOs, respectively.
Inserting the above mentioned Fourier ansatz into Eqs.~(\ref{Eq9})--(\ref{Eq11}), linearizing the
resultant system of equations, and sorting into equations for different Fourier modes, one obtains
the nonlinear dispersion relation \citep{Shukla07}
\begin{equation}
1 + \left(\frac{1}{\widetilde{D}_{+}}+\frac{1}{\widetilde{D}_{-}}\right) 
\left( 1+\frac{K^2}{D_L} \right)\frac{|A_0|^2}{2\gamma_0^3}=0,
\label{Eq12}
\end{equation}
where $\widetilde{D}_{\pm}=\pm 2\epsilon (\Omega- K U_g)-K^2$ and $D_L=1-\epsilon^2 + H_e^2K^4/4$.
One notes that $D_L=0$ yields the linear dispersion relation $\Omega^2=1+H_e^2K^4/4$ for
the EPOs in a dense quantum plasma \citep{Pines61}. For $H_e\rightarrow 0$ we recover
from (\ref{Eq12}) the nonlinear dispersion relation for relativistically large amplitude EM
waves in a classical electron plasma \citep{McKinstrie92}. The dispersion relation (\ref{Eq12})
governs stimulated Raman backward and forward scattering instabilities, as well as the modulational
instability. In the long wavelength limit $U_g \ll 1$, $\epsilon \approx 1$ one can use the
ansatz $\Omega= i\Gamma$, where the normalized (by $\omega_{pe}$) growth rate $\Gamma\ll 1$,
and obtain from Eq.~(\ref{Eq12}) the growth rate $\Gamma=(1/2)|K|\{(|A_0|^2/\gamma_0^3)[1+K^2/(1+H_e^2
K^4/4)]-K^2\}^{1/2}$ of the modulational instability. For $|K|<1$ and $H_e <1$, the linear growth rate
is only weakly depending on the quantum parameter $H_e$.

The quantum dispersion effects on nonlinearly coupled CPEM and EPOs can be studied by considering a
steady state structure moving with a constant speed $U_g$. Inserting the ansatz
$A_\perp= W(\xi)\exp(-i\Omega_e t)$, $\psi= P(\xi)\exp(ik_e x-i\omega_e t)$ and $ \phi = \phi(\xi)$ into
Eqs.~(\ref{Eq9})--(\ref{Eq11}), where $\xi=z-U_g t$, $k_e=U_g/H_e$ and $\omega_e = U_g^2/2H_e$,
and where $W(\xi)$ and $P(\xi)$ are real, one obtains from (\ref{Eq9})-(\ref{Eq11}) the coupled
system of equations \citep{Shukla07}
\begin{equation}
\frac{\partial^2 W}{\partial \xi^2}+\left (\lambda -\frac{P^2}{\gamma}+1\right)W=0,
\label{Eq13}
\end{equation}
\begin{equation}
\frac{H_e^2}{2}\frac{\partial^2 P}{\partial \xi^2}+(\phi-\gamma+1)P=0,
\label{Eq14}
\end{equation}
where $\gamma=(1+W^2)^{1/2}$, and
\begin{equation}
\frac{\partial^2 \phi}{\partial \xi^2}=P^2-1,
\label{Eq15}
\end{equation}
with the boundary conditions $W=\Phi=0$ and $P^2=1$ at $|\xi|=\infty$. In Eq. (\ref{Eq13}),
$\lambda = 2 \epsilon \Omega_e$ represents a nonlinear frequency shift of the CPEM wave. In the limit
$H_e\rightarrow 0$, one has from (\ref{Eq14}) $\phi= \gamma -1$, where $P \neq 0$, and one recovers
the classical (non-quantum) case of the relativistic solitary waves in a cold plasma \citep{Marburger75}.

The system of equations (\ref{Eq13})--(\ref{Eq15}) admits a Hamiltonian
\begin{equation}
\begin{split}
& Q_H =\frac{1}{2}\left(\frac{\partial W}{\partial\xi}\right)^2+\frac{H_e^2}{2}
\left(\frac{\partial P}{\partial\xi}\right)^2-\frac{1}{2}
\left(\frac{\partial\phi}{\partial\xi}\right)^2
\\
&+\frac{1}{2}(\lambda+1)W^2+P^2-\gamma P^2+\phi P^2-\phi=0,
\end{split}
\label{Eq16}
\end{equation}
where the boundary conditions $\partial/\partial\xi=0$, $W=\phi=0$ and $|P|=1$
at $|\xi|=\infty$ have been used.

\begin{figure}[htb]
\centering
\includegraphics[width=8cm]{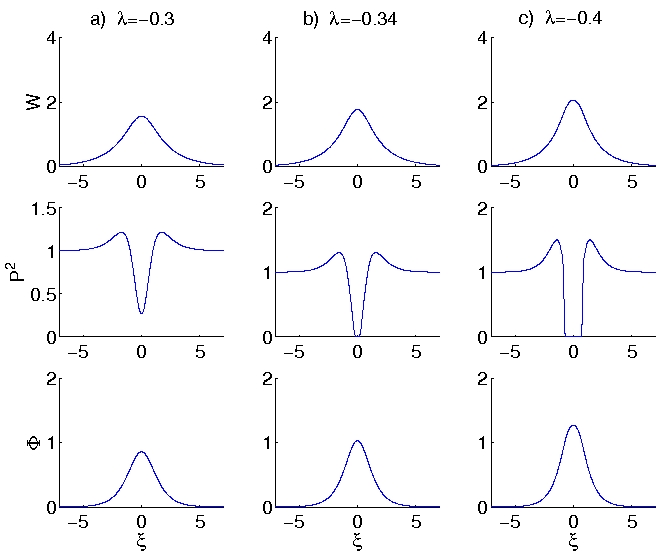}
\caption{The profiles of the CPEM vector potential $W$ (top row), the electron number density
$P^2$ (middle row) and the scalar potential $\Phi$ (bottom row) for $\lambda=-0.3$ (left column),
$\lambda=-0.34$ (middle column) and $\lambda=-0.4$ (right column), with $H_e=0.002$.
After \citet{Shukla07}.}
\label{Fig6}
\end{figure}

\begin{figure}[htb]
\centering
\includegraphics[width=8cm]{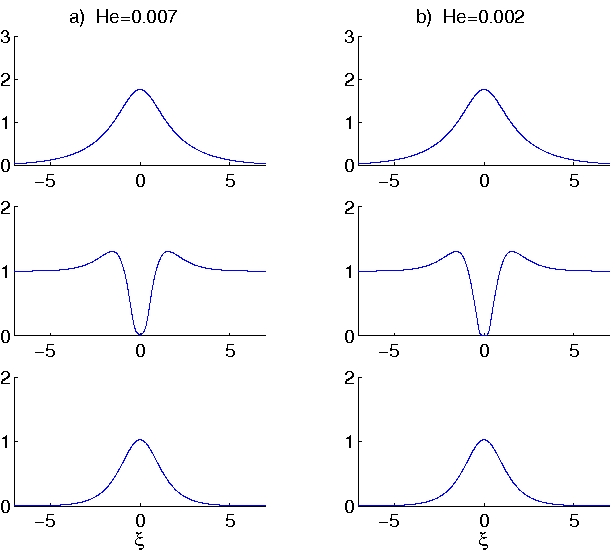}
\caption{The profiles of the CPEM vector potential $W$ (top row), the electron number density
$P^2$ (middle row) and the scalar potential $\Phi$ (bottom row) for $H_e=0.007$ (left column)
and $H_e=0.002$ (right column), with $\lambda=-0.34$. After \citet{Shukla07}.}
\label{Fig7}
\end{figure}

\begin{figure}[htb]
\centering
\includegraphics[width=8cm]{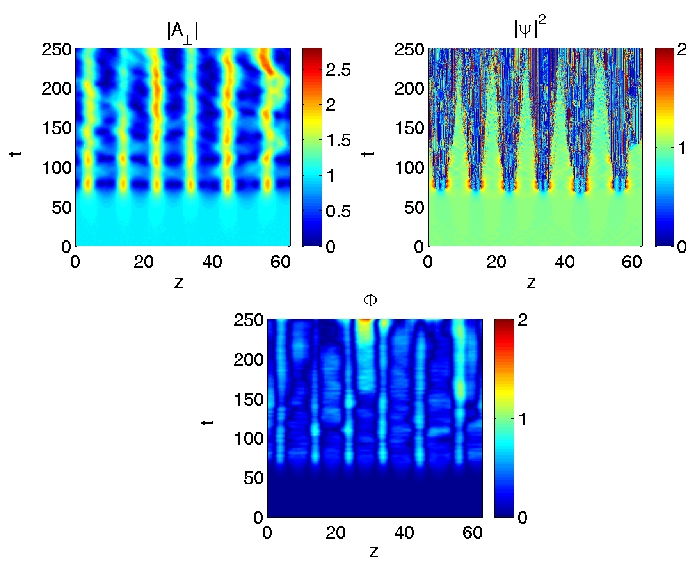}
\caption{The dynamics of the CPEM vector potential ${A}_\perp$ and the electron number
density $|\psi|^2$ (upper panels) and of the electrostatic potential $\Phi$ (lower panel)
for $H_e=0.002$. After \citet{Shukla07}.}
\label{Fig8}
\end{figure}

\begin{figure}[htb]
\centering
\includegraphics[width=8cm]{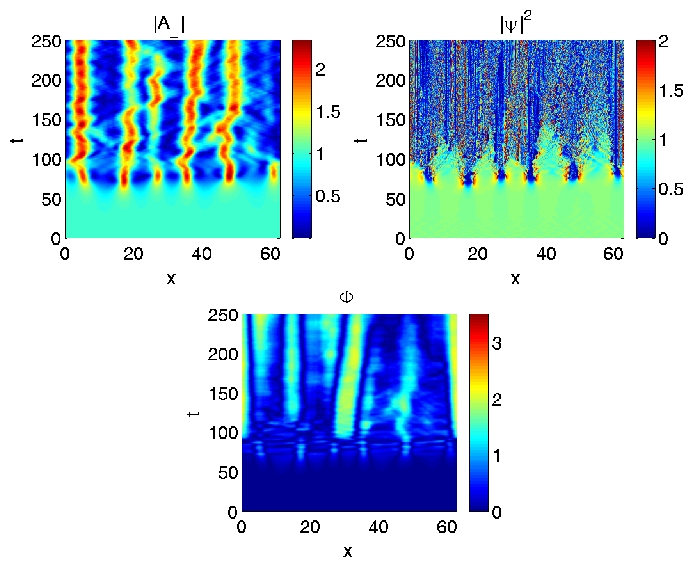}
\caption{ The dynamics of the CPEM vector potential ${A}_\perp$ and the electron number density $|\psi|^2$
(upper panels) and the electrostatic potential $\phi$ (lower panel) for $H_e=0.007$.  After \citet{Shukla07}.
}
\label{Fig9}
\end{figure}

Numerical solutions of the quasi-stationary system (\ref{Eq13})--(\ref{Eq15}) are presented
in Figs. \ref{Fig6} and \ref{Fig7}, while time-dependent solutions of Eqs.~(\ref{Eq9})--(\ref{Eq11})
are displayed in Figs. \ref{Fig8} and \ref{Fig9}.  Here parameters were used that are representative
of the next generation of laser-based plasma compression (LBPC) schemes \citep{Malkin07, Azechi06}.
The formula \citep{Shukla86} $eA_\perp/mc^2 =6 \times 10^{-10} \lambda_s \sqrt{I}$ will determine
the normalized vector potential, provided that the CPEM wavelength $\lambda_s$ (in microns) and
the CPEM wave intensity $I$ (in W/cm$^2$) are known. It is expected that in LBPC schemes, the
electron number density $n_0$ may reach $10^{27}$ cm$^{-3}$ and beyond, and the peak values of
$e A_\perp/mc^2$ may be in the range 1-2 (e.g. for focused EM pulses with  $\lambda_s \sim 0.15$ nm
and $I \sim 5 \times 10^{27}$ W/cm$^{2}$). For $\omega_{pe} =1.76 \times 10^{18}$ s$^{-1}$,
one has $\hbar \omega_{pe} =1.76 \times 10 ^{-9}$ erg and $H_e =0.002$, since $mc^2 =8.1 \times 10^{-7}$ erg.
The electron skin depth $\lambda_e \sim 1.7$ {\AA}. On the other hand, a higher value of $H_e=0.007$ is
achieved for $\omega_{pe} =5.64 \times 10^{18}$ s$^{-1}$. Thus, the numerical solutions below, based on
these two values of $H_e$, have focused on scenarios that are relevant for the next generation intense
laser-solid density plasma interaction experiments \citep{Malkin07}.

Figures \ref{Fig6} and \ref{Fig7} exhibit numerical solutions of Eqs. (\ref{Eq13})--(\ref{Eq15})
for $H_e=0.002$ and $H_e=0.007$.  The nonlinear boundary value problem was solved with the
boundary conditions $W=\phi=0$ and $P=1$ at the boundaries at $\xi=\pm 10$. The
solitary envelope pulse is composed of a single maximum of the localized vector potential $W$
and a local depletion of the electron density $P^2$, and a localized positive potential $\phi$
at the center of the solitary pulse. The latter has a continuous spectrum in $\lambda$, where
larger values of negative $\lambda$ are associated with larger amplitude solitary EM pulses.
At the center of the solitary EM pulse, the electron density is partially depleted, as in panels a)
of Fig. \ref{Fig6}, and for larger amplitudes of the EM waves one has a stronger depletion of the
electron density, as shown in panels b) and c) of Fig. \ref{Fig6}. For cases where the electron density
goes to almost zero in the classical case \citep{Marburger75}, one important quantum effect is that
the electrons can tunnel through the depleted density region. This is seen in Fig. \ref{Fig7}, where
the electron density remains nonzero for $H_e=0.007$ in panels a), while the density shrinks to zero
for $H_e=0.002$ in panel b).

Figures \ref{Fig8} and \ref{Fig9} depict numerical simulation results of Eqs.~(\ref{Eq9})--(\ref{Eq11})
for the long-wavelength limit characterized by $\omega_0\approx 1$ and $V_g\approx 0$. As initial conditions,
we used an EM pump with a constant amplitude $A_\perp=A_0=1$ and a uniform plasma density $\psi=1$,
together with a small amplitude noise (random numbers) of order $10^{-2}$ added to ${A}_\perp$
to give a seeding any instability.  The numerical results are displayed in Figs. \ref{Fig8}
and \ref{Fig9} for $H_e=0.002$ and $H_e=0.007$, respectively.  In both cases, we see an initial
linear growth phase and a wave collapse at $t\approx 70$, in which almost all the CPEM wave energy
is contracted into a few well separated large-amplitude, localized CPEM envelopes, associated with 
an almost complete depletion of the electron density
at the center of the CPEM wavepacket, and a large-amplitude positive electrostatic potential.
One can see that there is a more complex dynamics of
localized CPEM wavepackets for $H_e=0.007$, shown in Fig. \ref{Fig9}, in comparison with $H_e=0.002$,
shown in Fig. \ref{Fig8}, where the wavepackets are almost stationary when they are fully developed.

\section{Magnetized Quantum Plasmas}

Magnetized quantum plasmas occur in white dwarf stars and on the surface of magnetized
stars (e.g. magnetars) where degenerate electrons could be ultra-relativistic, but the ions are
in a non-degenerate state. How strong magnetic fields in dense stars come about is still a mystery,
although there are evidence of the strong magnetization of dense plasmas in astrophysical environments.
In dense magnetized plasmas, one has to account for the Lorentz force and the Landau quantization
effect \citep{Landau98a}, and develop the appropriate quantum Hall-magnetohydrodynamics (Q-HMHD)
equations starting from the Wigner-Maxwell equations. We stress, however, that the Q-HMHD equations
discussed here does not capture
the particular physics of the quantized Hall resistance $R_k = \hbar/ \nu e^2$ \citep{Klitzing1980}.
In semiconductors with 2D electrons, the latter is associated with the quantized electron  density
$n_q = \nu e B_0/\hbar c$ at high magnetic fields and low temperature, where $\nu$ is an integer, appearing
in the electron current ($- e n_q {\bf u}_d$) flowing through a conductor.
Here ${\bf u}_d =(c/B_0^2) {\bf E} \times {\bf B}_0$ is the cross-field electron drift associated with
the space charge electric field ${\bf E}$ that results from the motion of electrons by the Lorentz force.
The Ohm's law, in turn, determines the von Klitzing resistance, which is independent of the magnetic field.

\subsection{Landau Quantization}

In a strong magnetic field $\hat {\bf z} B_0$, where $\hat {\bf z}$ is the unit vector along the $z$-axis
in a Cartesian coordinate system, and $B_0$ the strength of the external magnetic field, the electron motion
in a plane perpendicular to the magnetic field direction is quantized \citep{Landau98b}. The electron
energy level is determined by the non-relativistic limit by the expression \citep{Landau98b,Tsintsadze2010}

\begin{equation}
{\cal E}_e^{l,\sigma} = \frac{p_z^2}{2m_e} + (2l + 1 + \sigma) \mu_B B_0,
\end{equation}
where $p_z$ is the electron momentum in the $z-$ direction, $l$ the orbital angular number ($l =0, 1,2)$,
and $\sigma =\pm 1$ represents the spin orientation. For $\sigma =-1$, we have from (88)

\begin{equation}
{\cal E}_e^{l} = \frac{p_z^2}{2m_e} + l \hbar \omega_{ce},
\end{equation}
where $\omega_{ce} =eB_0/m_ec$ is the electron gyrofrequency. Accordingly, the
Fermi-Dirac electron distribution is \citep{Tsintsadze2010}

\begin{equation}
F_D(p_z,l) \propto \frac{1}{1+ \exp \left[\left(E_z + l\hbar \omega_{ce} -\mu_e\right)/k_B T_e\right]},
\end{equation}
where $E_z =(m_e/2) v_z^2$ is the parallel (to $\hat {\bf z}$) kinetic energy of degenerate electrons.

Assuming that $|l\omega_{ce}-\mu_e| \gg k_B T_e$, one can approximate the Fermi-Dirac distribution function
by the Heaviside step function $H(\mu_e -{\cal E}_e^l)$, which equal $1$ for $\mu_e = E_{Fe} = k_B T_{Fe}=
(p_F^2/2m_e)^{1/2} > {\cal E}_e^l$ and zero for $E_{Fe} < {\cal E}_e^l$, where $p_F =m_e V_{TF}$.
The equilibrium electron number density is \citep{Tsintsadze2010}

\begin{equation}
n_e =\frac{p_F^3}{2\pi^2 \hbar^3}\left[\Gamma_B + \frac{2}{3}(1-\Gamma_B)^{3/2}\right],
\end{equation}
where $\Gamma_B =\hbar \omega_{ce}/k_B T_{Fe}$. The current carried by degenerate electrons in a
magnetized quantum plasma is $-e n_e {\bf u}_d$, which yields the plasma resistivity $R_s = en_ec/B_0$.

\subsection{ESOs and EM Waves}

In a magnetized quantum plasma, there are finite density perturbations associated with high-frequency
electrostatic electron-Bernstein (EB) waves and elliptically polarized EM waves (EP-EM waves)
that propagate across the magnetic field direction $\hat {\bf z}$. Furthermore, the CPEM wave
propagating along $\hat {\bf z}$ are not associated with any density perturbation.

\begin{figure}
\centering
\includegraphics[width=8.5cm]{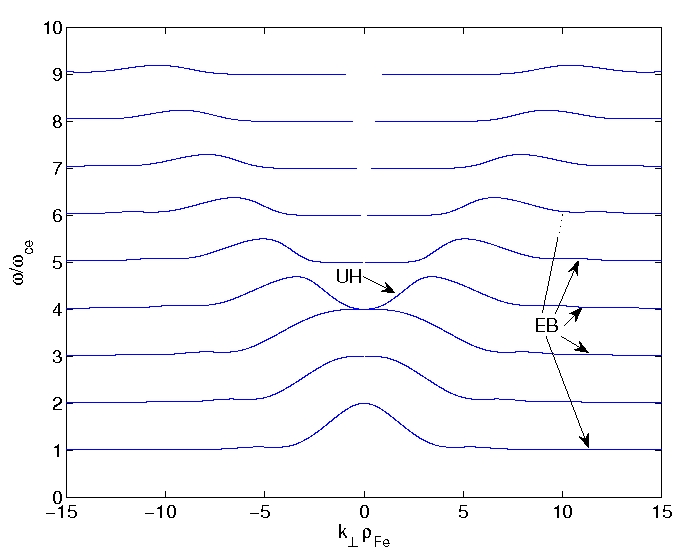}
\caption{Dispersion curves for EB waves in a Fermi-Dirac distributed plasma, showing
several EB modes and the UH cutoff.  After \citet{Eliasson08b}.}
\label{Fig_UH}
\end{figure}

The dispersion relation for the EB waves in a Fermi-Dirac distributed plasma is in the ultra-cold limit 
\citep{Eliasson08b}

\begin{equation}
1 + \frac{3\omega_{pe}^2}{\omega_{ce}^2 }
\int_0^ \pi d\theta \frac{\sin( \Omega \theta)\sin(\theta) \sin(\xi_e)
-\xi_e \cos(\xi_e)}{\xi_e^3} =0,
\end{equation}
where $\Omega =\omega/\omega_{ce}$, $\xi_e =(2k_\perp^2\rho_{Fe}^2)\cos(\theta/2)$, and $\rho_{Fe}
=V_{Fe}/\omega_{pe}$ is the gyroradius of degenerate electrons. Solutions of
Eq. (92) are plotted in Fig. \ref{Fig_UH} for the case $\omega_{UH}=4\omega_{ce}$,
where  $\omega_{UH} =(\omega_{pe}^2 + \omega_{ce}^2)^{1/2}$ is the upper-hybrid (UH) resonance frequency.
In the long wavelength limit (viz. $k_\perp^2 \rho_{Fe} \ll 1$), Eq. (92) yields
\begin{equation}
\omega^2 =\omega_{UH}^2 + \frac{3}{5}\frac{\omega_{pe}^2 k_\perp^2 V_{Fe}^2}{(\omega^2 -4 \omega_{ce}^2)},
\end{equation}
where $k_\perp$ is the perpendicular (to $\hat {\bf z}$ component of the propagation wave vector.
For $\omega \approx \omega_H$, Eq. (93) reveals that the propagating UH waves have positive (negative)
group dispersion in plasmas with $\omega_{pe} > (<) \sqrt{3}\omega_{ce}$.

Furthermore, the refractive index $N_x$ for the EP-EM waves propagating along the $x$ axis
(which is orthogonal to $\hat {\bf z}$) is \citep{Shukla2007}
\begin{equation}
\begin{split}
\!\!\!\!\! N_x &= \frac{k_x^2 c^2}{\omega^2 } 
\\
 &= 1- \frac{\omega_{pe}^2}
{\omega^2}- \frac{\omega_{pe}^2 \omega_{ce}^2[1+ \eta(\alpha) k_x^2\lambda_b^2]}
{\omega^2 [\omega^2 -\omega_{UH}^2 + k_x^2V_{Fe}^2(1+ k_x^2\lambda_q^2)]},
\end{split}
\end{equation}
where $k_x$ is the $x$ component of the propagation wave vector, $\lambda_q^2 = \hbar^2 /4m_e V_{Fe}^2$,
$\lambda_b =\sqrt{\hbar/2m_e \omega_{ce}}$, $\eta (\alpha) = 2 \tanh(\alpha)$, $\alpha=\mu_B B_0/k_B T_{Fe}$.
Several comments are in order. First, we note that the electron spin-$1/2$ effect
enhances the electron gyrofrequency by a factor $(1+\eta k_x^2\lambda_b^2)^{1/2}$ in the numerator of the
third term in the right-hand side of (94). Second, the quantum Bohm force produces a dispersion term $\hbar
k^4/4m_e^2$ in the denominator of the third term in (94). Third, in the limit of vanishing $\hbar$, Eq. (95)
correctly reproduces the EP-EM wave dispersion relation. Furthermore, Eq. (94) reveals that the cut-off
frequencies (at $k_x=0$) in dense magnetoplasmas are
\begin{equation}
\omega=\omega_\pm = \frac{1}{2} \left[(4 \omega_{pe}^2+\omega_{ce}^2)^{1/2}\pm \omega_{ce}\right],
\end{equation}
which are the same as the cutoffs of the X (upper sign) and Z (lower sign) mode waves
in a classical plasma \cite{Chen06}. Short wavelength electromagnetic propagation in magnetized
quantum plasmas, including quantum electrodynamic effects, has also been considered by
\citet{Lundin07}.

The vector representation of spinning quantum particles in the quantum theory was first introduced by
\citet{Takabayasi55} who developed the QHD involving the evolution of the quantum particle spin.
The idea of Takabayasi has been further elaborated by \citet{Brodin10} in the context of the spin
contribution to the ponderomotive force of the magnetic field-aligned CPEM waves in a quantum magnetoplasma.
In fact, by using the non-relativistic electron momentum equation \citep{Brodin10}
\begin{equation}
\begin{split}
& m_e \left(\frac{\partial}{\partial t} + {\bf u}_e \cdot \nabla\right){\bf u}_e 
\\
& = -e \left( {\bf E}
+ \frac{1}{c} {\bf u}_e \times {\bf B}\right) -\frac{g}{\hbar}\mu_B \nabla ({\bf B} \cdot {\bf s}),
\end{split}
\end{equation}
and the spin evolution equation
\begin{equation}
\left(\frac{\partial}{\partial t} + {\bf u}_e \cdot \nabla \right) {\bf s} = \frac{g \mu_B}{\hbar}
\left({\bf B} \times {\bf s}\right),
\end{equation}
together with  Amp\`ere's law and suitable Maxwell's equation (incorporating the electron magnetization
current, ${\bf J}_M = -(4 \pi/c)(g \mu_B/\hbar)\nabla \times (n_e \times {\bf s})$,
due to the electron $1/2-$ spin effect), where ${\bf s}$ is the spin angular momentum,
with its absolute value $|{\bf s}| = s_0 = \hbar/2$. The quantity $g= 2.0023192$ is the electron Gaunt factor
(sometimes called the $g$ factor or spectroscopic splitting factor). The value $g=2$ is predicted
from Dirac's relativistic theory of the electron, while the correction to this value comes from
the quantum electrodynamics \citep{Bransden00,Kittel96}.

\citet{Brodin10} derived the spin ponderomotive force $\hat {\bf z}F_{s}$ for
the CPEM wave, where

\begin{equation}
F_s =\mp \frac{g^2 \mu_B^2}{m_e^2 \hbar^2}\frac{s_0}{(\omega \pm \omega_g)}
\left[\frac{\partial}{\partial z} -\frac{k}{(\omega \pm \omega_g)}\frac{\partial}{\partial t}
\right]|{\bf B}_w|^2.
\end{equation}
Here $\omega_g =g \mu_B B_0/\hbar$ the spin-precession frequency, and ${\bf B}_w$ is the
CPEM wave magnetic field. The spin ponderomotive force comes from the averaging of the third
term in (99) over the CPEM wave period $2\pi/\omega$. The  CPEM wave frequency $\omega$ is
determined from the dispersion relation

\begin{equation}
\left[1 \mp \frac{\omega_\mu}{(\omega \pm \omega_g)}\right]N_z^2
= 1- \frac{\omega_{pe}^2}{\omega (\omega \pm \omega_{ce})},
\end{equation}
where $N_z = k_z c/\omega$, $k_z$ is the component of the wave vector ${\bf k}$ along the $z$ axis,
$\omega_\mu= g^2 s_0 /4 m_e \lambda_e^2$, $\lambda_e =c/\omega_{pe}$, and the $ + (-)$ represents the
left- (right-) hand circular polarization. The $\omega_\mu$-term in (100) is associated with the electron
spin evolution. It changes the dispersion properties of the magnetic field-aligned EM electron-cyclotron
waves in a quantum magnetoplasma.  Furthermore, the spin-ponderomotive force induces a strong
spin-polarization of a quantum magnetoplasma.

It should be noted that there is also a standard non-stationary ponderomotive force
($\hat {\bf z} F_e$) \citep{Karpman77} of the CPEM waves arising from the averaging of the
nonlinear Lorentz force term $-(e/m_ec) \hat {\bf z} \cdot({\bf u}_e \times {\bf B}_w)$ over
the CPEM wave period $2\pi/\omega$, where

\begin{equation}
F_e =-\frac{e^2}{2m_e^2 \omega (\omega\pm \omega_{ce})}\left[\frac{\partial}{\partial z}
\pm \frac{k_z\omega_{ce}}{\omega(\omega \pm \omega_{ce})}\frac{\partial}{\partial t}\right]
|{\bf E}_w|^2,
\end{equation}
where ${\bf E}_w = (\omega/k_z c) {\bf B}_w$ is the CPEM wave electric field.

\subsection{Q-HMHD Equations}

To a first approximation, the dynamics of low phase speed (in comparison with the speed of light
in vacuum) electromagnetic waves in dense magnetoplasmas is modeled by the Q-HMHD equations.
The latter include the inertialess electron momentum equation

\begin{equation}
0 = -e n_e \left( {\bf E} + \frac{1}{c} {\bf u}_e \times {\bf B} \right) - \nabla P_C,
\end{equation}
where the quantum Bohm and quantum spin forces are supposed to be unimportant on the characteristic
scalelength of present interest. The degenerate electrons are coupled with the non-degenerate ions through
the EM forces. The ion dynamics is governed by the ion continuity equation (43) and the momentum equation
\begin{equation}
m_i n_i \frac{d{\bf u}_i}{dt} =  n_i e \left({\bf E} + \frac{1}{c} {\bf u}_i \times {\bf B}\right),
\end{equation}
where $d/dt = (\partial/\partial t) + {\bf u}_i \cdot \nabla$. For the sake of simplicity, we have 
here assumed that $\tau_m \partial/\partial t \ll 1$ and $\partial {\bf u}_i/\partial t \gg 
(\eta/\rho_i)\nabla \cdot \nabla {\bf u}_i + \rho_i^{-1}(\xi +\eta/3) \nabla (\nabla \cdot {\bf u}_i)$.
The EM fields are given by Amp\`ere's law

\begin{equation}
\frac{\partial {\bf B}}{\partial t} = - c \nabla \times {\bf E},
\end{equation}
and Maxwell's equation
\begin{equation}
\nabla \times {\bf B} = \frac{4 \pi e}{c} (n_i {\bf u}_i -n_e {\bf u}_i)
+\frac{1}{c}\frac{\partial {\bf E}}{\partial t}.
\end{equation}

By using (102), we can eliminate the electric field ${\bf E}$ from (103), obtaining for a
quasi-neutral ($n_e = n_i =n$) quantum magnetoplasma

\begin{equation}
m_i n  \frac{d{\bf u}_i}{dt} = - \nabla P_C - \frac{1}{8\pi} \nabla {\bf B}^2
+ \frac{({\bf B}\cdot \nabla){\bf B}}{4\pi}  ,
\end{equation}
where we have used (104) without the displacement current (the last term on the right-hand side) 
for the low-phase speed (in comparison with $c$) EM wave phenomena. By using the electric field from 
(102), we can write (103) as

\begin{equation}
\frac{\partial {\bf B}}{\partial t} = \nabla \times ({\bf u}_i \times {\bf B})
- \frac{m_i c}{e} \frac{d{\bf u}_i}{dt},
\end{equation}

Equations (43), (105) and (106) are the desired Q-HMHD equations for studying the linear and nonlinear 
dispersive EM waves, as well as new aspects of 3D quantum fluid turbulence in a quantum magnetoplasma
with degenerate electrons having Chandrasekhar's pressure law.  However, when the Landau quantization effect
in a very strong magnetic field is accounted for, one can replace $P_C$ by the appropriate pressure
law \citep{Eliezer05}

\begin{equation}
\begin{split}
P_L =&\frac{4eB_0(2m_e)^{1/2}E_F^{3/2}}{3(2\pi)^2 \hbar^2 c}
\\
&\times
\left[1+ 2\sum_{l=1}^{l_m}\left(1-\frac{l\hbar \omega_{ce}}{k_B T_{Fe}}\right)^{3/2}\right],
\end{split}
\end{equation}
where the value of $l_m$ is fixed by the largest integer that satisfies $k_B T_{Fe} -l \hbar \omega_{ce} \leq 0$.

\section{Summary and Outlook}

In this Colloquium paper, we have described the essential physics of quantum plasmas with
degenerate electron fluids. We have reviewed the properties of quantum plasmas and quantum models 
that describe the salient features of linear and nonlinear ES and EM waves. Specifically, the focus 
of the present colloquium article has been on developing the model nonlinear equations that depict 
new features of nonlinear waves and quantum electron fluid turbulence at nanoscales. Numerical simulations 
of the nonlinear Schr\"odinger (NLS)-Poisson equations reveal quasi-stationary, localized structures in 
the form of one-dimensional electron density holes (dark solitons) and 2D quantum electron vortices. 
These localized quantum structures, which are associated with a local depletion of the electron density and
a positive electrostatic potential, arise due to a balance between the nonlinear and dispersion effects
involved in the dynamics of nonlinearly interacting EPOs. In 2D, there exist a class of quantum
electron vortices of different excited states (charge states). Furthermore, numerical simulations
also depict that the time-dependent NLS-Poisson equations exhibit stability of a dark soliton
in one-space dimension. In 2D the dark solitons of the first excited state are stable and the preferred 
nonlinear state is in the form of quantum vortex pairs of different polarities.  The one-dimensional 
dark soliton and singly charged 2D quantum vortices are thus long-lived nonlinear structures at nanoscales. 
Also presented are theoretical and computer simulation studies of nonlinearly coupled intense EM waves and 
EPOs in an unmagnetized quantum plasma. We have reported new classes of  stimulated scattering instabilities 
of EM waves and trapping of intense EM waves in a quantum electron density hole.

It should be noted that inclusion of non-degenerate ion dynamics gives rise to new features to
linear and nonlinear IPOs \citep{Haas03,Eliasson08a}. Furthermore, nonlinear equations governing
the coupling between the dispersive Langmuir and ion-acoustic waves, which are known as the quantum
Zakharov equations \citep{Garcia05,Misra08,Haas09,Simpson09}, admit periodic, quasi-periodic, chaotic
and hyper-chaotic states \citep{Misra08}, in addition to arresting the Langmuir wave
collapse \citep{Haas09,Simpson09} due to quantum dispersion effects. There may also emerge new aspects of
nonlinear EPOs and IPOs when the particle trapping \citep{Jovanovic07} in the strong wave potential is 
included. Here one has to obtain nonlinear solutions of non-stationary Wigner-Poisson equations, which 
might reveal a modified (by the electrostatic wave potential) Fermi-Dirac electron distribution function. 
Furthermore, there is a scope for studying the collective nonlinear response of correlated Coulomb 
electron systems at finite temperatures by means of kinetic theory concepts \citep{Domps97} to incorporate 
collisions and Green's function methods originally developed by \citet{Baym61}. We note that the 
Baym-Kadanoff approach has been used by \citet{Kwong00} to investigate the dielectric properties 
(viz. inverse dielectric function and dynamic structure factor) of linear EPOs in a correlated electron gas. 
Furthermore, the ion-ion dynamic structure factor, which contains a wealth of information about ions 
including structure and low-frequency collective modes in a dense quantum plasma, has been 
studied by \citet{Murillo10}.

The field of the nonlinear quantum plasma physics is vibrant, and its potential applications rest on our
complete understanding of numerous collective processes in compact astrophysical objects, as well as in the
next-generation of intense laser-solid density plasma experiments and in the plasma assisted nanotechnology
(e.g. quantum free electron laser devices, quantum-diodes, metallic nanostructures, nanowires, nanotubes, etc.).
However, nonlinear quantum models presented in this Colloquium paper have to be further improved and generalized 
by including the effects of the electron exchange interactions, strong electron-electron correlations,
equilibrium inhomogeneities of the magnetic field and the plasma density, as well as fully relativistic 
and Landau quantization effects in a nonuniform quantum magnetoplasma. We have also to understand the features 
of quantum oscillations of electrons and possible formation of bound states of electrons in the presence of 
an external magnetic field. For this purpose, we have to calculate the interaction potential among highly 
correlated electrons and use molecular dynamic simulations to demonstrate attraction among electrons due to 
collective wave-quantum particle interactions that give rise to Cooper's pairing of degenerate electrons. 
Cooper's pairing of electrons could possibly provide a scenario of superconducting behavior of a quantum plasma. 
Furthermore, 2D system composed of electron clusters at finite temperature exhibits Wigner Coulomb 
crystallization \citep{Egger99,Filinov01}. The latter has been investigated by means of Monte Carlo simulations 
based on a quantum Hamiltonian with parabolic confining and Coulomb interaction potentials. 
In a nonuniform quantum magnetoplasma, we have ES drift waves \citep{Shokri99,Ali07,Saleem08}
which  can drastically affect the cross-field electron transport. For applications to plasma assisted nano-technology
devices (e.g. nonlinear electrostatic and electromagnetic surface waves in metallic nanostructure-devices,
photonic band gap and x-ray optical systems, quantum X-ray free-electron laser systems), one must also study
nonlinear collective processes by including both the electron spin-$1/2$ and quantum electron tunneling effects
on an equal footing. Finally, the localization of coupled ES and EM waves due to nonlinear quantum effects 
in a nonuniform quantum magnetoplasma with an arbitrary electron pressure degeneracy should provide clues
to the origin of very intense X-rays \citep{Coe78} and gamma rays \citep{Hurley05} from both astrophysical
and laboratory plasmas.

\section*{Acknowledgments} This research was supported by the Deutsche Forschungsgemeinschaft through
the project SH21/3-1 of the Research Unit 1048, and by the Swedish Research Council (VR).

\bibliographystyle{apsrmp}

\end{document}